\documentclass[aps, prb, twocolumn,show pacs,show keys,superscript address,superscript reference,floatfix]{revtex4-2}
\usepackage[colorlinks=true,citecolor=blue,linkcolor=blue,urlcolor=blue]{hyperref}

\usepackage[sort&compress]{natbib}
\usepackage{graphicx,times}
\usepackage{color}
\usepackage{amssymb}

\usepackage{subeqnarray}

\usepackage{bbm}
\usepackage{bm}
\usepackage{diagbox}
\usepackage{makecell}
\usepackage{amsmath}

\newcommand{\ie}{{i.e.}}

\DeclareMathAlphabet\mathbfcal{OMS}{cmsy}{b}{n}
\newcommand{\ket}[1]{\ensuremath{\left|{#1}\right\rangle}}
\newcommand{\bra}[1]{\ensuremath{\left\langle{#1}\right|}}

\begin{document}

\title{Charge qubits in the USC regime for quantum state transfer}
\author{F. A. C{\'a}rdenas-L{\'o}pez}
\affiliation{International Center of Quantum Artificial Intelligence for Science and Technology~(QuArtist)\\
	and Physics Department, Shanghai University, 200444 Shanghai, China}
 \affiliation{Forschungszentrum J\"ulich, Institute of Quantum Control (PGI-8), D-52425 J\"ulich, Germany} 
\author{J. Yu}
\affiliation{International Center of Quantum Artificial Intelligence for Science and Technology~(QuArtist)\\
	and Physics Department, Shanghai University, 200444 Shanghai, China}
\author{C. K. Andersen}
\affiliation{QuTech and Kavli Institute of Nanoscience, Delft University of Technology, 2600 GA Delft, The Netherlands}
\author{E. Solano}
\affiliation{International Center of Quantum Artificial Intelligence for Science and Technology~(QuArtist)\\
	and Physics Department, Shanghai University, 200444 Shanghai, China}
\affiliation{Department of Physical Chemistry, University of the Basque Country UPV/EHU, Apartado 644, E-48080 Bilbao, Spain}
\affiliation{IKERBASQUE, Basque Foundation for Science, Plaza Euskadi, 5, 48009 Bilbao, Spain}
\affiliation{Kipu Quantum, Greifswalderstrasse 226, 10405 Berlin, Germany}

\author{A. Parra-Rodriguez}
\affiliation{Department of Physics, University of the Basque Country UPV/EHU, Apartado 644, E-48080 Bilbao, Spain}
\affiliation{Institut quantique \& Département de Physique, Université de Sherbrooke, Sherbrooke J1K 2R1 QC, Canada}

\begin{abstract}

We study the feasibility of reaching the ultrastrong (USC) and deep-strong coupling (DSC) regimes of light-matter interaction, in particular at resonance condition, with a superconducting charge qubit, also known as Cooper-Pair box (CPB). We numerically show that by shunting the charge qubit with a high-impedance LC-circuit, one can maximally reach both USC and DSC regimes exceeding the classical upper bound $|g|\leq \sqrt{\omega_q\omega_r}/2$ between two harmonic systems with frequencies $\omega_q$ and $\omega_r$. As an application, we propose a hybrid system consisting of two CPBs ultrastrongly coupled to an LC-oscillator as a mediator device that catalyzes a quantum state transfer protocol between a pair of transmon qubits, with all the parties subjected to local thermal noise. We demonstrate that the QST protocol maximizes its efficiency when the mediator operates in the USC regime, exhibiting comparable times with proposals relying on highly coherent and controllable mediators requiring state-preparation or post-selection. This work opens the door for studying light-matter interactions beyond the quantum Rabi model at extreme coupling strengths, providing a new building block for applications within quantum computation and quantum information processing.
\end{abstract}

\maketitle

\section{Introduction}
Macroscopic devices exhibiting quantum behavior have attracted much attention in the last three decades. The idea of having access to a system with tailored and tunable parameters paves the way for developing protocols to solve particular computational problems. Among the available quantum architectures, outstanding results~\cite{Arute2019} have been achieved with superconducting circuits~\cite{Vion2002,Clarke2008,Nori1,Krantz2019}, where one can study and engineer light-matter interactions in what is known as the field of circuit quantum electrodynamics (cQED)~\cite{Blais2004,Wallraff2004,Chiorescu2004,Schoelkopf2008,Blais2020}, in opposition to cavity quantum electrodynamics (CQED)~\cite{Walther2006,Hood2000,Raimond2001}. In this platform, one can design circuits containing Josephson junctions~\cite{	Josephson1962,Buttiker1987}, lumped elements that exhibit a wide range of anharmonicity, to mimic atomic energy spectra. These artificial atoms have been classified into three different types depending on the main degree of freedom~\cite{Yan2020}: charge qubits~\cite{Bouchiat1998,Nakamura1999,Koch2007,Schreier2008,Barends2013}, phase qubits~\cite{Martinis2002,Steffen2002,Ansmann2009}, and flux qubits~\cite{Orlando1999,Mooij1999,Shcherbakova2015}. On the other hand, LC oscillators and transmission line resonators encode (approximate) single, and multimode light degrees of freedom~\cite{Girvin2011,Nori3,Itoh1974,Goppl2008,Gely2017,ParraRodriguez2018}, respectively.

The fast development of cQED technology was triggered by the success of a system based on the charge qubit presenting coherent quantum interaction with a superconducting resonator~\cite{Blais2004,Wallraff2004}. This anharmonic system was made of a Josephson junction, with energy ($E_J$), in parallel with a capacitor, with dominating kinetic energy ($E_C\geq E_J$). However, this system proved to be very sensitive to charge fluctuations, and consequently, has shown very low coherence times. Later on, better performances were achieved with transmon qubits~\cite{Koch2007,Schreier2008}, a more harmonic version of the CPB where the potential Josephson energy dominates over the kinetic energy ($E_J\gg E_C$). Likewise, modified versions~\cite{Yan2016} of the original flux qubit, i.e., shunting the junctions with big capacitors in parallel, led to more harmonic and reproducible designs and are competing with transmon-like technology. Another leap in decay time performance was achieved with the fluxonium qubit~\cite{Manucharyan2009}, where their original lack of good coherence properties has been recently tackled~\cite{Nguyen2019}. More recent developments exploit multinodal configurations of linear and nonlinear elements~\cite{Smith2020,Kalashnikov2020,Gyenis2021} with potential improvements in both charge and flux noise protection.

These flexible technological accomplishments have encouraged the quest for engineering previously unobserved coupling regimes in natural systems such as the ultrastrong (USC)~\cite{Nori2,FornDiaz2019}, and deep-strong (DSC) coupling regime~\cite{Casanova2010,Yoshihara2017} in cQED. In the minimal USC regime scheme, the coupling strength between a qubit (matter) and a harmonic oscillator mode (light) is comparable to or larger than the subsystem energies of the two coupled systems. Here, the complete quantum Rabi model (QRM) \cite{Rabi1936,Rabi1937,Braak2011} is necessary to unveil the correct non-classical physics~\cite{Ballester2012,FornDiaz2010,Beaudoin2011}, e.g., such as in superradiant phase transitions~\cite{Nataf2010, Viehmann2011}. This is due to the fact that the rotating wave approximation (RWA) leading to the Jaynes-Cummings model is no longer valid~\cite{Casanova2010,Braak2011,Rossatto2017}. Distinct spectroscopic features of this system have been experimentally probed with superconducting flux \cite{FornDiaz2010,Niemczyk2010,FornDiaz2017,Yoshihara2017} and transmon qubits~\cite{BosmanPRB2017,Bosman2017,Kuzmin2019} embedded in cQED setups, as well as with quantum dots~\cite{Scarlono2021} coupled to high impedance resonators. In addition, dynamical features of this regime regime have been observed in a decelerated frame~\cite{Braumueller2017}. Furthermore, it is well known that the discrete parity symmetry and an anharmonic energy spectrum that the QRM presents provide a set of resources for quantum information and quantum simulations \cite{Nataf2011,Romero2012,Kyaw2015,Felicetti2015,KyawPRB2015,Wang2016,Arriagada2018,Garziano2015,KockumPRA2017,Kockum2017,Stassi2017,Stassi2020}.

All different proposals for building superconducting qubits coupled to harmonic oscillators (light photons) suggest the existence of a natural trade-off between anharmonicity on the qubit, the strength that the coupling between them could have, and the noise experienced in the device. In charge qubits, considerable anharmonicity increases the sensitivity to the voltage (charge) fluctuations that controls it \cite{Bouchiat1998,Nakamura1999}. For flux qubits, the anharmonic energy spectrum produces high sensitivity to external magnetic flux \cite{Orlando1999,Shcherbakova2015}. In these devices, attenuating the anharmonicity may lead to the enhancement of their coherence times. Nevertheless, the coupling strength that can be achieved with harmonic degrees of freedom decreases \cite{Devoret2007,ManucharyanJPA2017,Gely2018}. In fact, it is already known that the coupling parameter $g$ between two harmonic oscillators with frequencies $\omega_q$ and $\omega_r$ is bounded by $g\leq \sqrt{\omega_q\omega_r}/2$ \cite{BosmanPRB2017,Jaako2016,FornDiaz2019}. It has been however shown that such restriction does not exist when we consider a harmonic oscillator coupled to an anharmonic system, e.g., a 
superconducting charge qubit~\cite{Jaako2016}. Moreover, it has been shown that the topology of the connection between the defined anharmonic and harmonic degrees of freedom plays a crucial role in this bound~\cite{ManucharyanJPA2017}.

In this article, we investigate theoretically the spectroscopic and dynamical properties of a charge superconducting qubit ($\omega_q$) coupled to an LC oscillator ($\omega_r$) in its most general (mixed series/parallel) configuration beyond the USC regime. Here, we aim to find experimentally feasible sets of circuital parameters such that we have both resonance and USC/DSC regime, readily $g\sim \{\omega_r,\omega_q\}$ and $\omega_r\simeq\omega_q$, thus beating the previously mentioned purely harmonic limit. As an application, we prove that an extension of the proposed USC qubit-oscillator system, now with two qubits and an oscillator, can be used as a new quantum information building block, i.e., the mediator of a quantum state transfer (QST) protocol, working in the presence of a thermal environment. We show that maximal fidelity is attained when the mediator operates in the USC regime and that it exhibits competitive times in comparison with gates implemented with coherent and controllable couplers.

The paper is structured as follows. In Sec.~\ref{sec2} we derive the Hamiltonian of one and two charge qubits coupled to the LC oscillator. Then we study numerically and analytically the coupling limit of the model we proposed, and the conditions needed for the different parties to be both on resonance, and in the USC/DSC coupling regime. Furthermore, we provide feasible circuit parameters for the experimental implementation of both the single and the two-CPB models to achieve the resonant USC/DSC regime. In Sec.~\ref{sec3} we study the properties of the two-qubit model, and use it as the incoherent mediator for performing QST. Finally, we summarize the main conclusions of this work in Sec.~\ref{sec5}.

\section{Coupling ratio of the light-matter interaction}
\label{sec2}
In recent decades, there has been an increasing interest to explore and push the light-matter coupling to its limits. Constraint by the fine-structure constant $\alpha\approx 1/137$, coupling parameters in CQED are bounded. On the other hand, engineered quantum systems in circuit QED have not been found to present such limitation~\cite{Devoret2007}, and experimental and theoretical efforts to find the true coupling limit between subsystems have been prolific~\cite{FornDiaz2010,Niemczyk2010,Nataf2010,Viehmann2011,FornDiaz2017,FornDiaz2019,Bosman2017,Yoshihara2017,ManucharyanJPA2017,Hita2022}.

In this section, we review and extend the works of~\cite{ManucharyanJPA2017,FornDiaz2019,Jaako2016}, and study the coupling limit of the light-matter interaction with a charge qubit coupled to an LC oscillator on and off resonance. We show that the existing bound on the coupling versus local frequencies ratio of two harmonic systems is no longer valid here, i.e., there is no fundamental limit for an anharmonic system, like the charge qubit, that prevents it from reaching the non-perturbative USC regime~\cite{Rossatto2017} (npUSC) \emph{on} resonance. Finally, we extend this result to a two-qubit and oscillator circuit, which we later use in the QST protocol.
 
\subsection{Hamiltonian description}
\begin{figure}[!ht]
\includegraphics[width=0.9\linewidth]{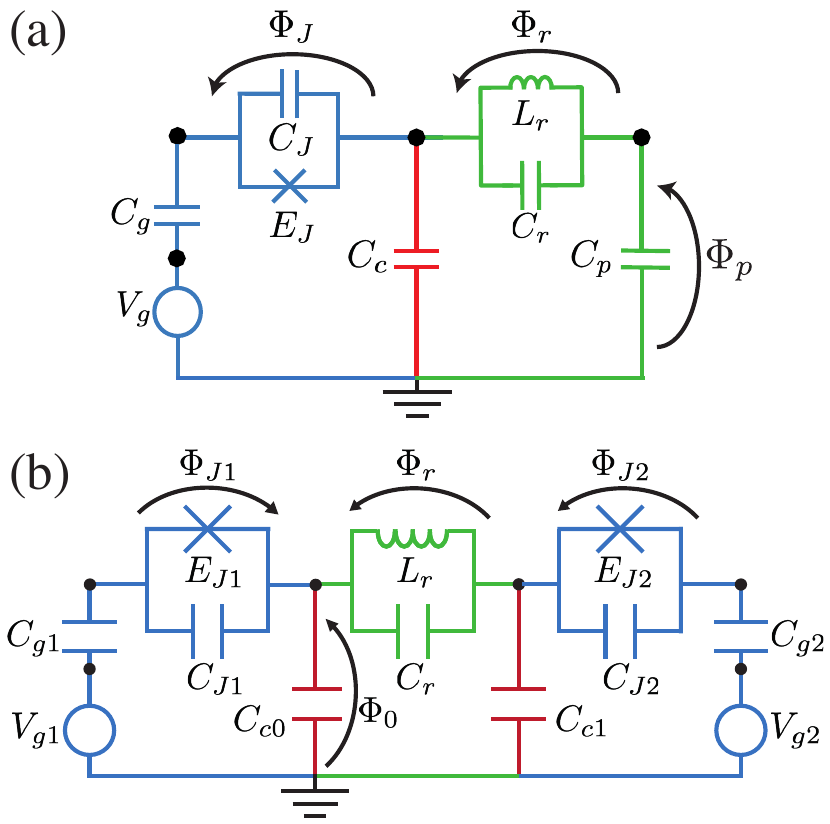}
\caption{(a) Illustration for a Cooper-pair box capacitively coupled to an LC oscillator. The CPB corresponds to a Josephson junction with Josephson energy $E_{J}$, and capacitance $C_{J}$ is biased by an external gate source $V_{g}$ through a capacitor $C_{g}$. The LC oscillator corresponds to two capacitances $C_{r}$, and $C_{p}$ in series with an inductor $L_r$. We describe the system with two independent flux node variables, $\Phi_{J}$ and $\Phi_{r}$. (b) Extended symmetrical circuit with two Josephson junctions and an LC oscillator. Both $\Phi_0$ and $\Phi_p$ node variables can be eliminated in the analysis.}
\label{model}
\end{figure}

Let us study the circuit depicted in Fig.~\ref{model}(a), consisting of a  Cooper-pair box (CPB) generically coupled with an LC oscillator. The CPB is formed by a Josephson junction with Josephson energy $E_{J}$ and capacitance $C_{J}$. This junction couples to a superconducting island providing Cooper-pairs from a voltage source $V_{g}$ through a gate capacitance $C_{g}$. The oscillator is formed by a lumped inductor $L_{r}$ and capacitor $C_{r}$. There are ground connections from the oscillator nodes through capacitors $C_{c}$ and $C_{p}$. Notice that this circuit recovers pure series and parallel coupling configurations in the limit of $C_p\rightarrow\infty$, and $C_c\rightarrow 0$, respectively~\cite{ManucharyanJPA2017}.

We describe the electrical circuit in terms of the flux node variables related to the voltage drops as $\Phi=\int_{-\infty}^{t}V(t')dt'$, leading to the following Lagrangian 
\begin{eqnarray}
\label{Eqn1}
L_{1}=\frac{1}{2}\dot{\vec{\Phi}}^{T}\mathbfcal{C}\dot{\vec{\Phi}}-\tilde{V}_g\dot{\vec{\Phi}}^{T}\mathbfcal{C}_{v}-\frac{\Phi_r^2}{2L_r}+E_J\cos{\big(\varphi_J\big)},
\end{eqnarray}
here, $\vec{\Phi}=\{\Phi_{J},\Phi_{r}\}$ correspond to the flux vector describing the CPB and the LC oscillator, and $\varphi_J=\Phi_J/\varphi_0$ is the superconducting phase, where $\varphi_{0}=\hbar/2e$ is the reduced magnetic flux quantum, and $2e$ is the charge of a Cooper-pair. Notice that we have eliminated the flux coordinate $\Phi_p$ since it corresponds to a passive node (for more details see Appendix~\ref{appendixA}). $\mathbfcal{C}$ is the capacitance matrix of the circuit, and $\mathbfcal{C}_{v}$ is the gate capacitance vector. We obtain the circuit Hamiltonian by computing the canonical conjugate momenta $Q_{\ell}=\partial L_1/\partial[\dot{\Phi}_{\ell}]$, and applying the Legendre transformation leading to 
\begin{align}\nonumber
H_{1} = & \frac{1}{2\tilde{C}_J}(Q_J-2en_g)^2-E_J\cos{(\varphi_J)}\\
&+\frac{Q_r^2}{2\tilde{C}_r} + \frac{\Phi_r^2}{2L_r}+\frac{Q_r(Q_J-2 e n_g)}{C_{Jr}},
\label{Eq02}
\end{align}
where $Q_J$ is the CPB charge, and $Q_r$ is the oscillator charge. Furthermore, $\tilde{C}_J$, $\tilde{C}_r$, and ${C}_{Jr}$ correspond to the effective Josephson, oscillator, and coupling capacitances, respectively. $n_{g}$ is the effective gate-charge number, which depends on the collective voltage bias $V_g$, and we set it throughout the whole article to the sweet spot of the CPB, \ie $n_{g}=0.5$. The quantization of the Hamiltonian in Eq.~(\ref{Eq02}) is done by promoting conjugated variables to quantum operators satisfying the commutation relations $[e^{i\hat{\varphi}_{J}},\hat{n}_{J}]=e^{i\hat{\varphi}_{J}}$, and $[\hat{\Phi}_{r},\hat{Q}_{r}]=i\hbar$, where $\hat{n}_J=\hat{Q}_J/2e$ quantifies the excess of Cooper-pairs on the CPB, $e^{i\hat{\varphi}_J}=\sum_{n_J}\ket{n_{J}+1}\bra{n_{J}}$ changes by one the number of Cooper-pairs, and we write the oscillator operators in terms of annihilation and creation operators $\hat{\Phi}_{r} = i\sqrt{\hbar \omega_{r} L_r/2}(a-a^\dagger)$, and $\hat{Q}_{r} = \sqrt{\hbar \omega_{r} \tilde{C}_r/2}(a+a^\dagger)$, satisfying the commutation relation $[a,a^{\dag}]=1$, and the effective oscillator frequency $\omega_r=1/\sqrt{\tilde{C}_r L_r}$. Finally, we obtain the quantum Hamiltonian as follows
\begin{align}\nonumber
H_1 = &\,4E_C (\hat{n}_J-n_g)^2 - E_J\cos{(\hat{\varphi}_J)}+ \hbar\omega_r a^\dag a \\
& + \hbar g(\hat{n}_J-n_g)(a + a^\dag),
\label{Eq04}
\end{align}
where $E_C={e^2}/2\tilde{C}_J$ is the charge energy of the CPB, and 
\begin{align}
 g &= {\omega}_{r}\gamma \sqrt{\frac{\tilde{Z}_{r}}{2R_Q}},
\end{align}
is the coupling strength between the CPB and the LC oscillator, where $\tilde{Z}_{r} = \sqrt{L_r/\tilde{C}_r}$ is the effective impedance of the oscillator. $R_Q = \hbar/(2e)^2$ is the reduced quantum resistance, and $\gamma=\tilde{C}_r/C_{Jr}$ is the ratio between the dressed resonator capacitance and the coupling capacitance. It must be remarked that the coupling operator can be expressed as proportional to $\hat{n}_J(a+a^\dag)$ if we displace the oscillator with $D(\alpha)=e^{\alpha a^\dag - \alpha^* a}$, i.e., $\tilde{H}_1=D^\dag\left( \frac{n_g g}{\omega_r} \right)H_1 D\left( \frac{n_g g}{\omega_r} \right)$, at the expense of modifying the pure Josephson kinetic energy with an additional term $2 n_g g/\omega_r \hat{n}_J$. Be aware that a second displacement in the Josephson subsector of the phase-space would not cure the issue but rather return back to $H_1$. We remind the reader that only in the limit of two coupled harmonic oscillators, the coupling operator is insensitive to the charge offset, as it happens for the low energy sector of the transmon circuit~\cite{Koch2007}.

An analogous analysis can be performed to a system composed of two CPBs coupled to the same LC oscillator as shown in Fig.~\ref{model}(b). Following the same procedure as before, we obtain the Hamiltonian
\begin{align}
H_{2} = & \sum_{\ell=1}^2 H^{\rm{cpb}}_{\ell}+ \hbar \omega_{r}a^\dagger a+\hbar\sum^2_{\ell=1}  g_\ell(\hat{n}_{J\ell}-n_{g\ell})(a+a^\dagger)\nonumber\\
& + \hbar g_{12}(\hat{n}_{J1}-n_{g1}) (\hat{n}_{J2}-n_{g2}), \label{eq:H_2CPB_LC_osc}                
\end{align}
where $H^{\rm{cpb}}_{\ell}=4E_{C\ell} (\hat{n}_{J\ell}-n_{g\ell})^2 - E_{J\ell}\cos(\hat{\varphi}_{J\ell})$ is the Hamiltonian of the $\ell$th CPB, and again, we assume that both CPBs work on the sweet spot \ie $n_{g\ell}=0.5$. Here, $g_\ell$ is the coupling strength between the $\ell$-th CPB and LC oscillator, and $g_{12}$ stands for the direct coupling parameter between the two CPBs, defined as 
\begin{align}
 g_{\ell}&=\omega_r\gamma_\ell\sqrt{\frac{\tilde{Z}_r}{2R_Q}},\quad
 g_{12}= -\frac{1}{R_Q C_{12}},
\end{align}
with $\gamma_\ell=\tilde{C}_r/C_{\ell r}$, and $\ell\in\{1,2\}$. Moreover, $\tilde{C}_r$ is the dressed resonator capacitance,  $C_{\ell r}$ is the coupling capacitance between the $\ell$th CPB and the oscillator, $C_{12}$ is the dressed coupling capacitance between the CPBs. Notice that even though both CPBs have no direct coupling between them, an effective capacitive coupling arises, through the capacitance matrix inversion (see Appendix~\ref{appendix_series_CPB_Ham} for more details). Furthermore, we have discarded the constant term $\mathcal{O}(n_{g\ell}^2)$ which does not modify the equations of motion.

\subsection{Limit of the light-matter coupling ratio}
Now that we have obtained the exact Hamiltonians for the circuits above, let us study the limit of the light-matter coupling ratio. As previously mentioned, we consider the CPB working on the sweet spot, where its low-lying energy spectrum can in certain conditions be truncated to an effective two-level system, with transition frequency $\hbar\omega_{q}\approx E_J$. The coupling operator written in the truncated eigenstate basis of the CPB Hamiltonian reads $(\hat{n}_{J}-n_g) = \sigma^x/2$, where $\sigma^x$ is the first Pauli matrix. Under this approximation, the Hamiltonian in Eq.~(\ref{Eq04}) becomes the QRM
\begin{equation}
H_1\approx\frac{\hbar \omega_{q}}{2}\sigma^{z} +\hbar\omega_{r} a^{\dag}a +  \hbar \tilde{g}\sigma^x(a + a^\dagger),
\label{Eq07}
\end{equation}
\begin{figure}[t!]
\includegraphics[width=1\linewidth]{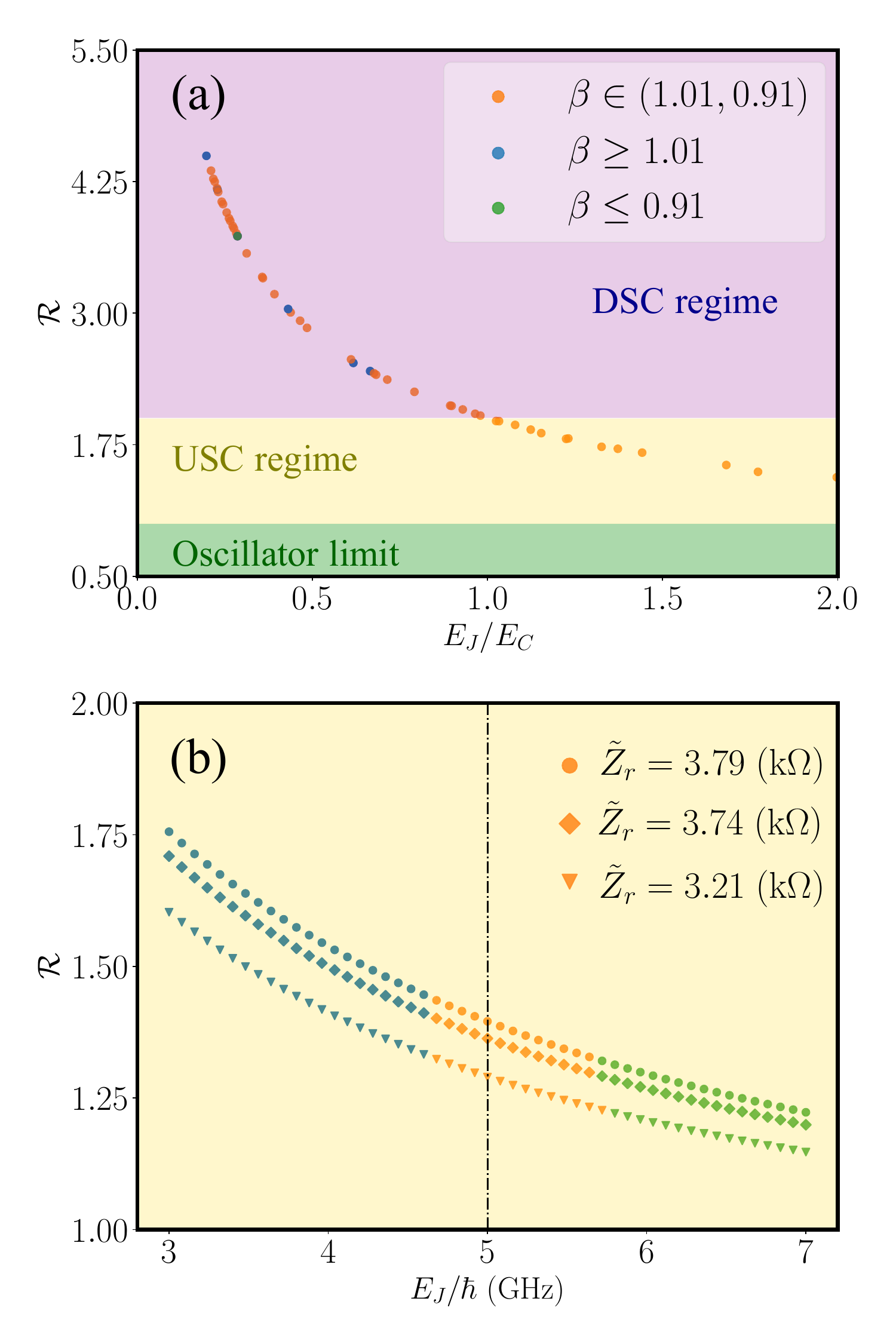}
\caption{Ratio $\mathcal{R}$ as a function of: (a) $E_J/E_C$, and (b) $E_{J}$ for resonators with different impedance $\tilde{Z}_{r}$ but identical resonator frequency $\omega_{r}$. In both cases, the  results (circuit parameter data-sets from which to compute the dots) were obtained with the Python minimizer routine, whereas the solid lines, corresponding to Eq.~({\ref{Eq09}}), are drawn to guide the eye. We classify the dots by shape and color according to the ratio $\beta=\omega_{q}/\omega_{r}$.}
\label{ratios} 
\end{figure}
where $\tilde{g}=g/2$. In recent years, the exploration of the fundamental limits of the coupling between an anharmonic system (matter) and a single-mode resonator (light) has drawn great attention~\cite{Devoret2007,Jaako2016,ManucharyanJPA2017,FornDiaz2019}. Within the two-level approximation of the matter degree of freedom, the ratio~\cite{Nataf2010,Jaako2016,BosmanPRB2017}
\begin{eqnarray}
\mathcal{R}\equiv\frac{2\tilde{g}}{\sqrt{\omega_{r}\omega_{q}}}=\frac{1}{C_{Jr}}\sqrt{\frac{\tilde{C}_{r}}{2 R_Q E_J }}.
\label{Eq09}
\end{eqnarray}
is a compact figure of merit, gauging the strength of the coupling with respect to the characteristic energies of the qubit and the harmonic part. A generalization of such quantity can be done beyond for higher-energy transitions in the anharmonic subsystem. For the case at hand, it can be readily appreciated that the coupling ratio is inversely proportional to the Josephson energy $\sqrt{E_J}$ of the CPB, and can, in principle, be arbitrarily increased~\cite{Jaako2016}. This formula generalizes those defined in \cite{FornDiaz2019,ManucharyanJPA2017}, which are proposed for resonance condition, i.e.,  $\omega_r=\omega_q$,
\begin{eqnarray}
\mathcal{R}_{\text{res}}\equiv \frac{2\tilde{g}}{\omega_r}\bigg|_{\omega_r=\omega_q}=\gamma x,
\label{Eq10}
\end{eqnarray}
where $x=\sqrt{\tilde{Z}_r/(2R_Q)}$ is a dimensionless parameter, and the capacitance ratio is
$\gamma=C_g C_{p}/(C_{g}(C_{c}+C_{p}) + C_{J}(C_{c}+C_{p}+C_{g}))$. 

Observe that by considering $C_g, C_p\gg C_c, C_J$, the ratio $\mathcal{R}_{\text{res}}\approx x$, i.e., it increases proportional to the square root of the resonator impedance~\cite{ManucharyanJPA2017}. In other words, there is no fundamental restraint preventing us from reaching the USC/DSC regime when the anharmonic subsystem works as a CPB. Conversely, it is possible to recover an upper bound~\cite{FornDiaz2019,BosmanPRB2017,Jaako2016} for such ratio in the harmonic (transmon) limit of the nonlinear subsystem (Josephson variable), i.e., replacing the pure Josephson term with a linear inductor ($L_J$) in Fig.~\ref{model}(a), leading to
 \begin{align}
 g_o&= \sqrt{\frac{\omega_q\omega_r \tilde{C}_J \tilde{C}_r}{4C^2_{J_r}}}\leq \frac{\sqrt{\omega_r \omega_q}}{2} \, .
\label{Eq12}
\end{align}
Here, $g_o$ is the effective coupling strength between the harmonic systems, and $\omega_q=1/\sqrt{\tilde{C}_J L_J}$ is the frequency of the transmon-oscillator, see Appendix~\ref{appendix_two_oscillator_model} for more details.

To gain further insight into these ratios, we aim to find the optimal set of circuital parameters that allows us to exceed the {\it classical} upper bound set by $\mathcal{R}$. For this purpose, we propose the use of an objective function that maximizes the coupling strength $\tilde{g}$ constrained to $\omega_{q}=\omega_{r}$ (resonance condition)
\begin{align}
\mathit{F}_1=1-\frac{ \tilde{g}\left|\bra{e}\hat{n}_J-n_{g}\ket{g}\right|}{\Delta_+}\cdot\left[1-\frac{2\left|\Delta_-\right|}{\Delta_+}\right],
\label{Eq13}
\end{align}
where $\omega_q$ is the CPB frequency, and $\bra{e}\hat{n}_J-n_{g}\ket{g}$ corresponds to the matrix element of the displaced charge operator. The numerical optimization algorithm truncates the Hilbert space of the CPB to $N_{\text{CPB}}=5$ charge states. In addition, $\Delta_{\pm}=\omega_q\pm\omega_r$ corresponds to the detuning between the CPB and the resonator frequencies. We minimise over the space parameter in Table~\ref{Tab01} for the one CPB case, and Table~\ref{Tab02} for two CPBs case (with an accordingly modified objective function), respectively, where the circuital parameters were explored in the value ranges $C=(0.11,550)~{\rm{fF}}$ for the capacitances, $L=(100,600)~{\rm{nH}}$ for the inductances, and  $E_{J}=2\pi(6,11)~{\rm{GHz}}$ for the Josephson energies~\cite{Wendin2017,Blais2020}. For the single CPB case, we assume no constraints between parameters, whereas for the two CPBs we consider that the grounding capacitors $C_{c\ell}$ are identical during the optimization subroutine. See below in the following subsection further details about the physical realization of circuits with those parameters.

In Fig.~\ref{ratios}(a), we plot $\mathcal{R}$ as a function of the ratio $E_J/E_C$ color-classified by a resonance ratio ($\beta=\omega_q/\omega_r$), fixing all circuit parameters except the Josephson energy.  Naturally, on approaching the CPB regime ($E_{J}\sim E_{C}$), we can reach values of $\mathcal{R}>1$ (see Fig.~\ref{ratios}(a)), beating the oscillator (transmon) limit. Moreover, in Fig.~\ref{ratios}(b), $\mathcal{R}$ is plotted as a function of $E_{J}$ for different values of the resonator impedance $\tilde{Z}_{r}$, keeping fixed the frequency $\omega_{r}$. For instance, by considering $\tilde{C}_{r}\rightarrow\tilde{C}_{r}/\mu$, and $L_{r}\rightarrow \mu L_{r}$, we obtain a new impedance  smaller than the previous one $\tilde{Z}_{r}\rightarrow \mu\tilde{Z}_{r}$, while the frequency remains the same. Points (squares, bullets, and crosses) in Fig.~\ref{ratios}(b) correspond to the values obtained through optimization, with the color classifying their resonance condition $\beta$ as in the previous figure, while solid lines are computed from Eq.~(\ref{Eq09}) and a data-set point to guide the eye. This figure highlights the known fact~\cite{ManucharyanJPA2017,FornDiaz2019,Devoret2007} that the larger the resonator's impedance, the larger the coupling ratio $\mathcal{R}$ is, consolidating the result obtained in Eq.~(\ref{Eq10}) i.e., $\mathcal{R}\propto\sqrt{\tilde{Z}_{r}/2R_{Q}}$.

\subsection{Choice of Parameters}
Minimizing the cost function $\mathit{F}_1$ in Eq.~(\ref{Eq13}), we find circuital parameters for experimental implementation of the devices depicted in Fig.~\ref{model} such that their frequencies and coupling strengths beat the upper classical limit $2\tilde{g}\leq \sqrt{\omega_{q}\omega_{r}}$, while imposing the resonance condition. We summarize these values and their corresponding Hamiltonian parameters/ratios in Table~\ref{Tab01} for the single CPB coupled to the LC oscillator. We find that the explored parameters are consistent with previous cQED experiments~\cite{Wendin2017,Pechenezhskiy2020}, i.e., within the 5-11~GHz frequency range. Notice that, we must have a high-impedance oscillator in order to obtain a large coupling strength, which can be achieved either by building an array of Josephson junctions having a large effective inductance (values of $L\leq2.5$\,$\mu$H have already been engineered \cite{Pechenezhskiy2020,Andersen2016,Stockklauser2017}), or by increasing the kinetic inductance of a superconducting film~\cite{Grunhaupt2019}.

\begin{table}[h]
\renewcommand{\arraystretch}{1.2}
\begin{tabular}{|l|l|lllll}
\cline{1-4}
\multicolumn{2}{|c|}{Parameters} & \multicolumn{2}{c|}{Results}                                     &  &  &  \\ \cline{1-4}
$C_g$~[fF]         & 212        & \multicolumn{1}{l|}{$\omega_r/2\pi$~(GHz)}     & \multicolumn{1}{l|}{5.95} &  &  &  \\ \cline{1-4}
$C_J$~[fF]         & 3.60        & \multicolumn{1}{l|}{$\omega_q/2\pi$~(GHz)}     & \multicolumn{1}{l|}{5.95} &  &  &  \\ \cline{1-4}
$C_c$~[fF]         & 0.14        & \multicolumn{1}{l|}{$E_J/E_C$}           & \multicolumn{1}{l|}{1.41} &  &  &  \\ \cline{1-4}
$C_r$~[fF]         & 0.81        & \multicolumn{1}{l|}{$\tilde{Z}_r$~(k$\Omega$)} & \multicolumn{1}{l|}{6.05} &  &  &  \\ \cline{1-4}
$C_p$~[fF]        & 180       & \multicolumn{1}{l|}{$\tilde{g}/2\pi$~(GHz)} & \multicolumn{1}{l|}{5.00} &  &  &  \\ \cline{1-4}
$L_r$~[nH]      & 162        & \multicolumn{1}{l|}{$\mathcal{R}$}        & \multicolumn{1}{l|}{1.68} &  &  &  \\ \cline{1-4}
$E_J/h$~(GHz)          & 6.00        &                                      &                           &  &  &  \\ \cline{1-2}
\end{tabular}
\caption{Optimal values for the circuit parameters and the corresponding Hamiltonian parameters/ratios of the one-CPB model depicted in Fig.~\ref{model}(a), where the circuit values are obtained by minimizing the objective function $\mathit{F}_1$ in Eq. (\ref{Eq13}).} 
\label{Tab01}
\end{table}

\begin{table}[h]
\renewcommand{\arraystretch}{1.2}
\begin{tabular}{ll|l|l|lll}
\cline{1-4}
\multicolumn{2}{|c|}{Parameters}      & \multicolumn{2}{c|}{Results} &  &  &  \\ \cline{1-4}
\multicolumn{1}{|l|}{$C_{g1}$~[fF]}    & 100 & $\omega_r/2\pi$~(GHz)         & 10.9   &  &  &  \\ \cline{1-4}
\multicolumn{1}{|l|}{$C_{J1}$~[fF]}    & 5.00 & $\omega_{q,1}/2\pi$~(GHz)       & 10.8   &  &  &  \\ \cline{1-4}
\multicolumn{1}{|l|}{$C_{g2}$~[fF]}    & 100 & $\omega_{q,2}/2\pi$~(GHz)       & 10.9   &  &  &  \\ \cline{1-4}
\multicolumn{1}{|l|}{$C_{J2}$~[fF]}    & 1.00 & $E_{J1}/E_{C1}$             & 1.99   &  &  &  \\ \cline{1-4}
\multicolumn{1}{|l|}{$C_c$~[fF]}     & 20 & $E_{J2}/E_{C2}$             & 1.63   &  &  &  \\ \cline{1-4}
\multicolumn{1}{|l|}{$C_r$~[fF]}     & 0.70 & $\tilde{Z}_r$~(k$\Omega$)     & 7.61  &  &  &  \\ \cline{1-4}
\multicolumn{1}{|l|}{$L_r$~[nH]}   & 700 & $\tilde{g}_1/2\pi$~(GHz)    & 4.71   &  &  &  \\ \cline{1-4}
\multicolumn{1}{|l|}{$E_{J1}/h$~(GHz)} & 11.0 & $\tilde{g}_2/2\pi$~(GHz)    & 5.73   &  &  &  \\ \cline{1-4}
\multicolumn{1}{|l|}{$E_{J2}/h$~(GHz)} & 11.0 & $\tilde{g}_{12}/2\pi$~(GHz)   & 12.8   &  &  &  \\ \cline{1-4}
                               &      & $\mathcal{R}_1$           & 2.25   &  &  &  \\ \cline{3-4}
                               &      & $\mathcal{R}_2$           & 2.25  &  &  &  \\ \cline{3-4}
\end{tabular}
\caption{Optimal values for the circuit parameters and the corresponding Hamiltonian parameters/ratios of the two-CPB model depicted in Fig.~\ref{model}(b), where the circuit values are obtained by minimizing the objective function $\mathit{F}_2$ in Eq.~(\ref{F2}). We have considered two asymmetric CPBs and equal grounding capacitors $C_{c0}=C_{c1}=C_c$ in the optimization subroutine.} 
\label{Tab02}
\end{table}
We extend our study for two-CPBs coupled to the LC oscillator (depicted in Fig.~\ref{model}(b)). Here, it is convenient to define the analogous coupling ratio to Eq.~(\ref{Eq09}) for the $\ell$th CPB with frequency $\omega_{q,\ell}$ as follows (see Appendix \ref{appendix_series_CPB_Ham} for details)
\begin{eqnarray}
\mathcal{R_\ell}\equiv\frac{2\tilde{g}_\ell}{\sqrt{\omega_{r}\omega_{q,\ell}}}=\frac{1}{C_{{\ell }r}}\sqrt{\frac{\tilde{C}_{r}}{2 R_Q E_{J\ell} }}.
\label{Eq14}
\end{eqnarray}
As we are interested in comparing the coupling strength between the qubits with the LC oscillator, we constrain the optimization problem to suppress the direct qubit-qubit interaction i.e., $g_{12}\ll \tilde{g}_\ell$, with the new objective function
\begin{align}
\mathit{F}_2=1-f_{12}\prod^2_{\ell=1}\frac{ \tilde{g}_\ell\left|\bra{e_\ell}\hat{n}_{J\ell}-n_{g\ell}\ket{g_\ell}\right|}{\Delta^\ell_+}\cdot\left[1-\frac{2\left|\Delta^\ell_-\right|}{\Delta^\ell_+}\right],
\label{F2}
\end{align}
here, $f_{12}=1-{g_{12}}/({\omega_{q,1}+\omega_{q,2}})$ constrains the qubit-qubit coupling, and $\Delta^\ell_{\pm}=\omega_{q,\ell}\pm\omega_r$. Finally, $\ket{g_\ell}$, $\ket{e_\ell}$ are the ground and first excited states of the $\ell$-th CPB, respectively. In Table~\ref{Tab02}, we summarize the optimal circuit values and their corresponding Hamiltonian parameters for the aforementioned situation. The optimization routine gives us a slight asymmetry in the total capacitance in parallel with the junction, irrelevant for the qubit frequencies, which are well approximated by the Josephson energies in the CPB regime, but naturally providing different coupling strengths $\tilde{g}_{\ell}$.

In summary, we have numerically shown that a (or more) nonlinear system(s) coupled to a high-impedance LC circuit can beat the pure oscillators' (classical) limit $2\tilde{g}\leq \sqrt{\omega_{q}\omega_{r}}$ by engineering suitable parameters. Having obtained and analyzed the Hamiltonian for the two CPBs USC-coupled to an oscillator, let us study the feasibility of using this compound unit as an incoherent mediator~\cite{Cardenas2017} to perform a quantum state transfer protocol. 

\section{Incoherent coupler for quantum state transfer}
\label{sec3}
\begin{figure}[t]
\centering
\includegraphics[width=0.9\linewidth]{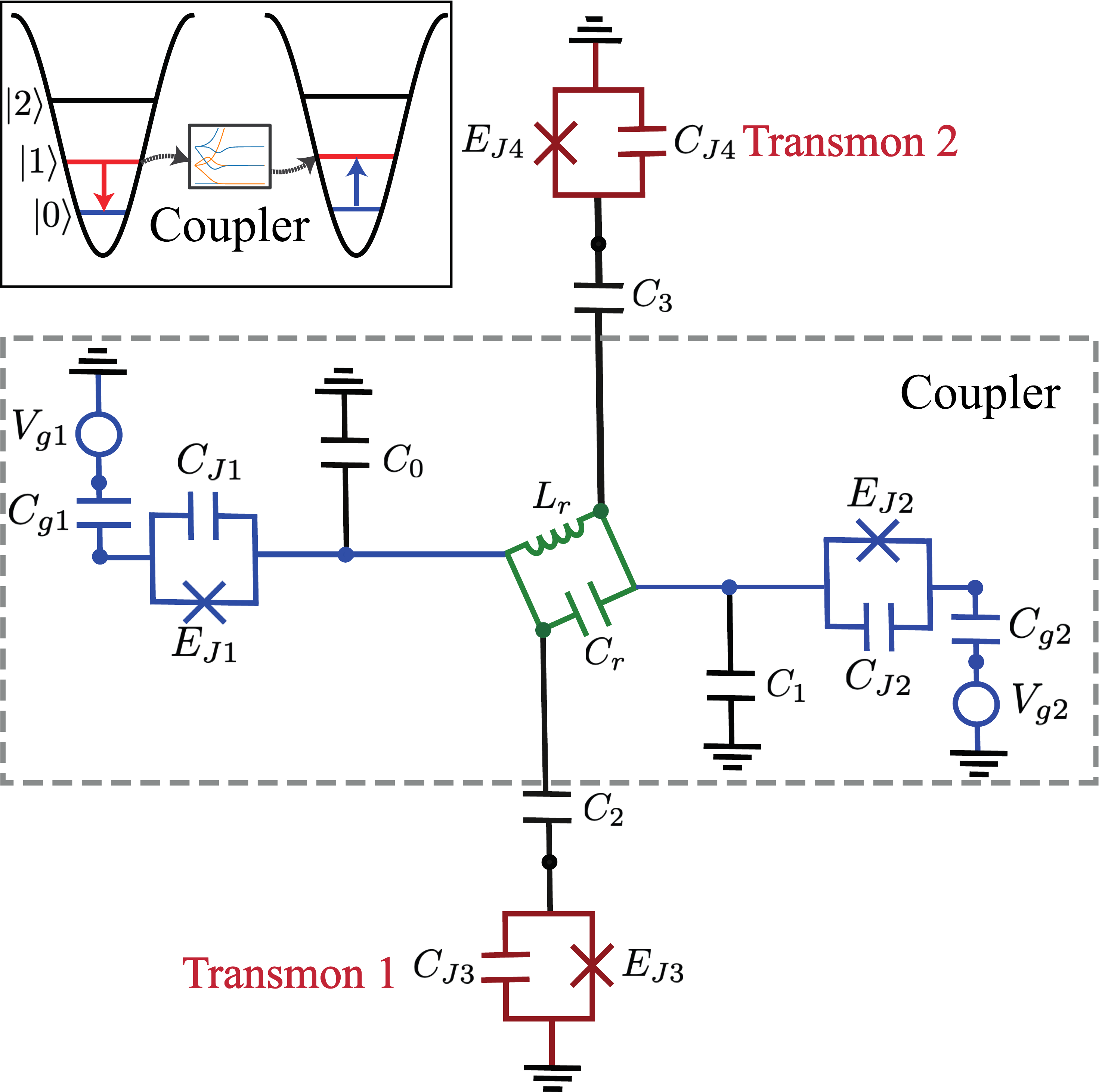}
\caption{Illustration of the QST protocol. An LC oscillator (green box) capacitively coupled to a pair of CPBs (blue boxes), and connected with grounded capacitors $C_{0}$, and $C_{1}$. This subsystem constitutes our mediator. We weakly couple two additional transmons (red boxes) to the LC oscillator, through capacitors $C_2$ and $C_3$. The illustration in the upper left corner is the schematic QST diagram.}
\label{qst}
\end{figure}
Among the several different tasks required to perform quantum processing, the coherent and fast controllable interaction between two subsystems is one of the most  crucial~\cite{Yan2018, Collodo2020}. However, a successful information exchange often relies on having a highly controllable and coherent third-party quantum system that acts as a mediator/coupler. A radically different approach is to design a protocol which is insensitive to the state of the mediator~\cite{Cardenas2017}, with the advantage that the coupler does not require preparation or post-selection~\cite{Collodo2020,google_sup,google_2,coupler2020}.

In this section, we follow the latter approach by Cardenas {\it et al.}~\cite{Cardenas2017} and consider a theoretical mediator described by the Hamiltonian in Eq. (\ref{eq:H_2CPB_LC_osc}), approximating the CPBs by two-level systems ($H_{\rm{coupler}}\approx H_2$) capacitively coupled to two transmons with strength $\lambda_1$ and $\lambda_2$. Since the CPB and the transmon have positive and negative anharmonicity, respectively, there is a choice of parameters for which the two transmons only couple to the oscillator in the one-excitation manifold~\cite{Ku2020}, thus obtaining the Hamiltonian of the QST protocol
\begin{align}
H&=H_{\rm{coupler}}+\sum_{m=1}^{2}\frac{\omega_{01,m}}{2}\tau_{m}^{z}+\sum_{m=1}^2\lambda_m\tau_{m}^{x} (a+a^\dagger),
\label{eqH}
\end{align}
where $\omega_{01,m}$, $\tau_{m}^{x}$, and $\tau_{m}^{z}$ correspond to the first frequency transition and the first and third Pauli matrices describing the $m$-th transmon, respectively.

\begin{figure}\includegraphics[width=0.95\linewidth]{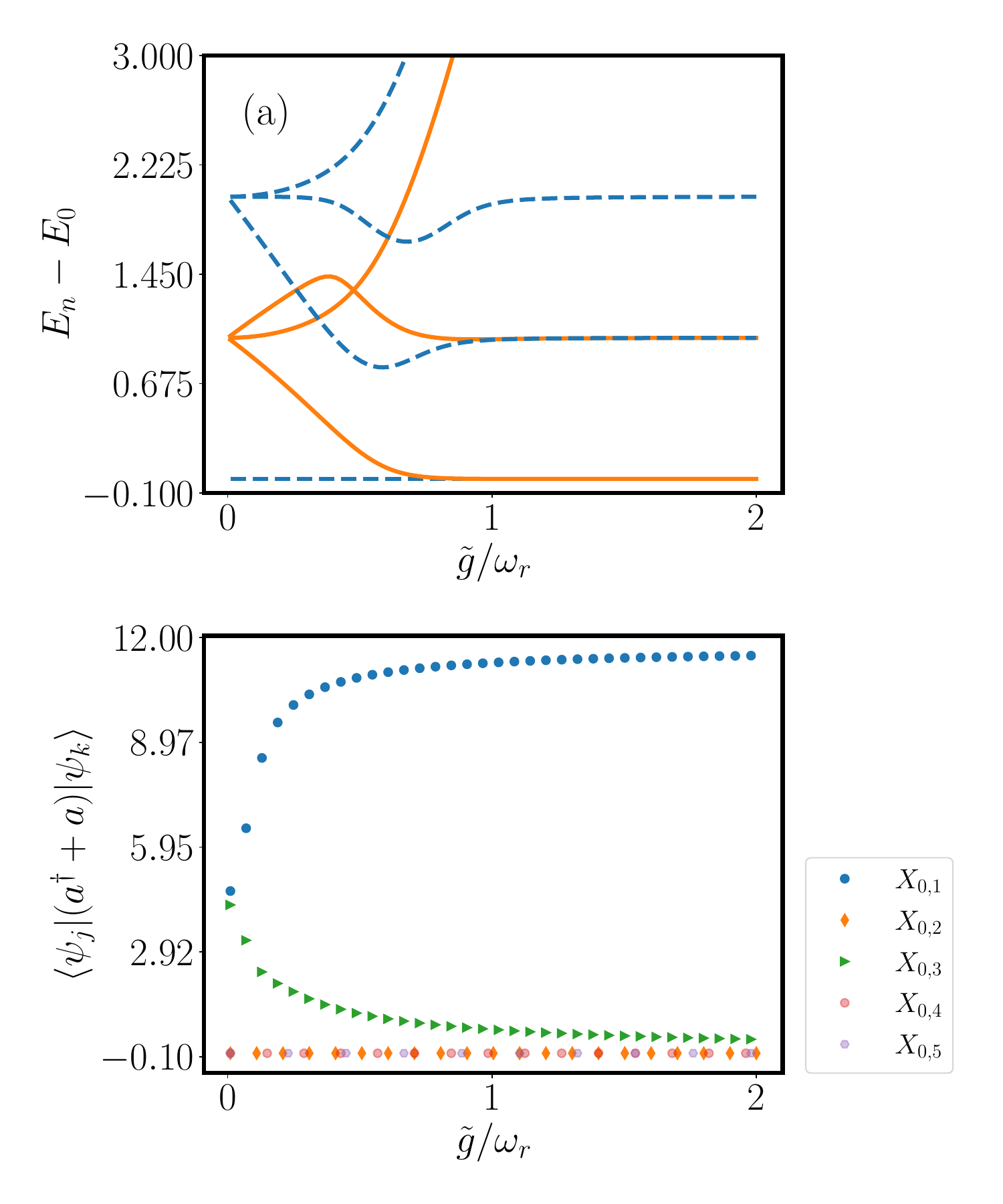}
\caption{Matrix element $\bra{\psi_j}a+a^{\dag}\ket{\psi_k}$ for different coupling ratios (a) $\tilde{g}_{1(2)}/{\omega}_r=0.3$, and (b) $\tilde{g}_{1(2)}/{\omega}_r=0.5$, where $\ket{\psi_{j(k)}}$ is the corresponding eigenstate of the mediator Hamiltonian in Eq.~(\ref{eq:H_2CPB_LC_osc}). The numerical calculations have been performed with $\omega_{q,1(2)}=\omega_{r}$.}
\label{spectrum} 
\end{figure}
The Hamiltonian $H_{\rm{coupler}}$ corresponds to the two-qubit Rabi model~\cite{Chilingaryan2013} that has an internal $\mathbb{Z}_{2}$ symmetry described by the parity operator $\mathcal{P} = \sigma_{1}^{z}\otimes\sigma_{2}^{z}\otimes \exp(i\pi a^\dag a)$ that divides the energy spectrum into two orthogonal subspaces that satisfy $\mathcal{P}\ket{\psi_{q}}=p\ket{\psi_{q}}$ ($p=\pm 1$), where $\ket{\psi_{q}}$ is the $q$~th eigenstate of the USC Hamiltonian, i.e., $H_{\rm{coupler}}=\sum_{q}E_{q}\ket{\psi_{q}}\bra{\psi_{q}}$, (see Fig~\ref{spectrum}{(b)}). It is now straightforward to demonstrate that these manifolds have specific selection rules over the operator $a^{\dag}+a$, i.e., $X_{pq}\propto\bra{\psi_{j}}(a^{\dag}+a)\ket{\psi_{j'}}\neq 0$ if $p\neq p'$~\cite{PDiaz2016}.\par
\begin{figure*}[t!]
	\includegraphics[width=0.9\linewidth]{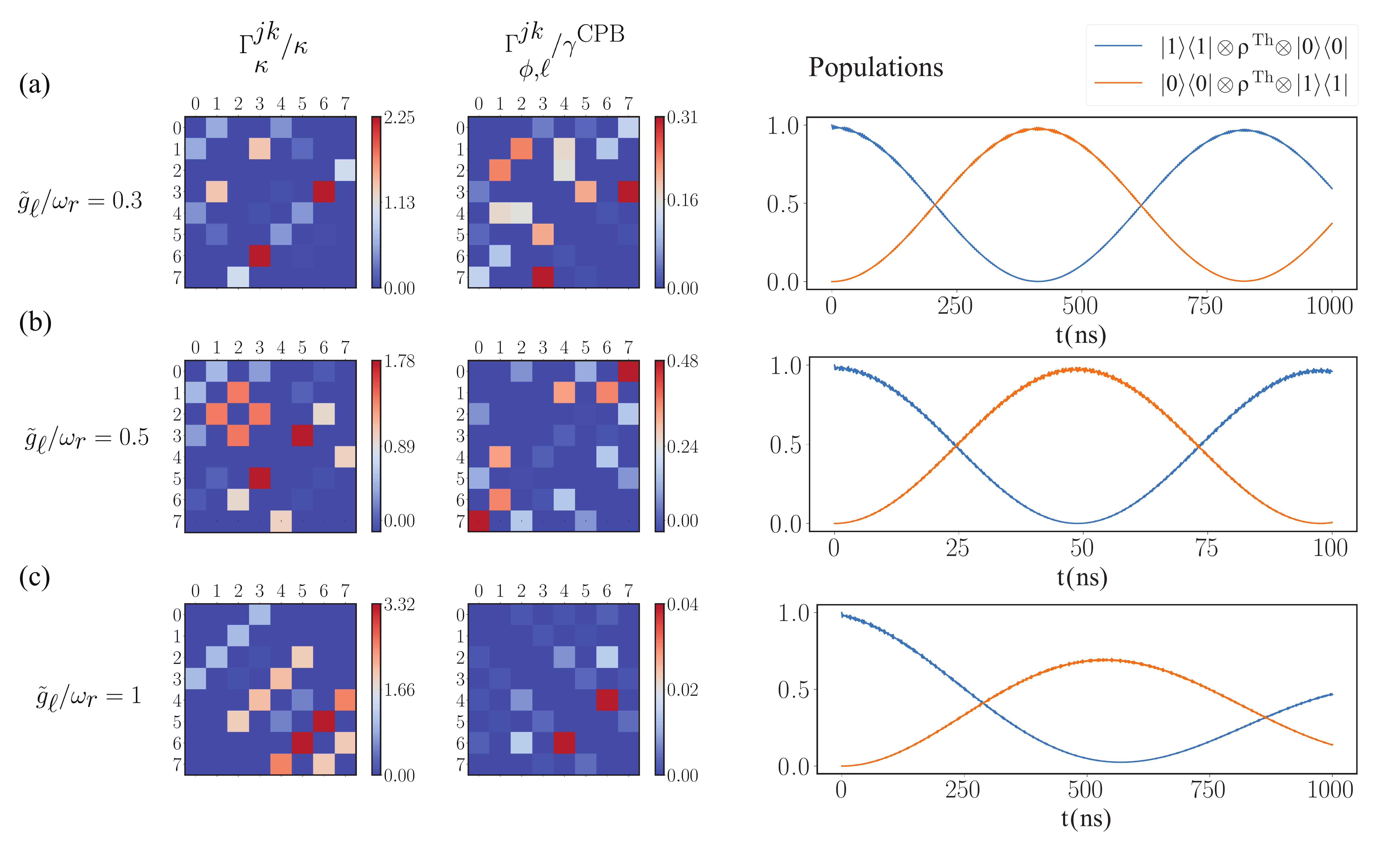}
	\caption{Matrix representation of the dressed decay rates (in units of the bare ones) $\Gamma^{jk}_\kappa$,$\Gamma^{jk}_{\gamma,\ell}$  ($\kappa$, $\gamma$) for different coupling strength (a) $\tilde{g}_{\ell}/\omega_{r}=0.3$, (b) $\tilde{g}_{\ell}/\omega_{r}=0.5$, and (c) $\tilde{g}_{\ell}/\omega_{r}=1$, respectively. Population inversion between the states $\rho_{1}=\ket{1}\bra{1}\otimes\rho^{{\rm{Th}}}\otimes\ket{0}\bra{0}$ (blue), and the state $\rho_{2}=\ket{0}\bra{0}\otimes\rho^{\rm{Th}}\otimes\ket{1}\bra{1}$ (orange), for a thermal state at ${T}=50~\rm{mK}$, calculated from the master equation in Eq. (\ref{Eq17}), with two-level transmons, and coupling parameter (a) $\tilde{g}=0.3\omega_{r}$, (b) $\tilde{g}=0.5\omega_r$, and (c) $\tilde{g}=\omega_r$. The above calculations have been performed using parameters $\omega_{q,{1(2)}}=\omega_r=2\pi\times8.13\textrm{(GHz)}$, and $\lambda_{1(2)}=0.02 \omega_r$.}
	\label{matrix_population} 
\end{figure*}

Let us now explore the performance of this system as a mediator for incoherent quantum state transfer when it operates in the perturbative USC (pUSC) regime and beyond \cite{Rossatto2017}. The protocol starts by selecting the first energy transition of the transmons on resonance with a forbidden energy transition of the mediator quadrature  $X=a+a^{\dag}$. Fig.~\ref{spectrum}{(b)} shows the transition matrix elements of the operator $X_{pq}$ as a function of $\Tilde{g}/\omega_{r}$ for the ground and first excited state, respectively. Here, we will focus on the transition matrix elements for three values of the coupling strength corresponding to the pUSC ($\Tilde{g}/\omega_{r}=0.3$), npUSC ($\Tilde{g}/\omega_{r}=0.5$) and DSC regime ($\Tilde{g}/\omega_{r}=1$), respectively. The forbidden energy transition corresponds to the $\ket{\psi_{0}}\leftrightarrow\ket{\psi_{3}}$ at $\Tilde{g}/\omega_{r}=0.3$, whereas for the npUSC and DSC regimes the forbidden transition is the $\ket{\psi_{0}}\leftrightarrow\ket{\psi_{2}}$.

We calculate the system dynamics considering losses mechanism, including energy relaxation and depolarizing channels. In the USC/DSC regime, the weak-coupling master equation formalism used in earlier quantum optics is not valid, because the dissipative dynamics strongly depends on the interaction between the subsystems. The corrected method~\cite{Beaudoin2011} requires to write the system-environment Hamiltonian in a diagonal basis for the USC/DSC system, leading to two principal physical consequences. Firstly, the decay rates depend on the energy transition of each level. Secondly, the coupling operator  $\sigma^{z}$ is not diagonal in the dressed basis. Therefore, the depolarizing channel has an additional component producing energy relaxation on the system. We encompass all these effects into the following master equation in the Lindblad form~\cite{Beaudoin2011, Settineri2018, Kryszewski2008} (see Appendix~\ref{appendixD} for more details)
\begin{align}\nonumber
\frac{d{\rho}(t)}{dt} = & -{i}[H,{\rho}(t) ]+\sum^{2}_{m=1}\left[\gamma_m \mathcal{D}[\tau^x_m]{\rho}(t)+\gamma_{\phi_m} \mathcal{D}[\tau^z_m]{\rho}(t)\right]\\\nonumber
&+\sum_{j,k>j}\left[\Gamma^{jk}_{\uparrow{\rm{eff}}}\mathcal{D}[\sigma_{jk}]\rho(t)+\Gamma^{jk}_{\downarrow{\rm{eff}}}\mathcal{D}[\sigma_{kj}]\rho(t)\right]\\
&+\sum_{\ell=1}^2\sum_{j}\Gamma_{\phi,\ell}^{jj}\mathcal{D}[\sigma_{jj}]{\rho}(t),
\label{Eq17}
\end{align}
where the Hamiltonian $H$ is given by Eq. (\ref{eqH}), $\sigma_{jk}=\ket{\psi_j}\bra{\psi_k}$ corresponds to an operator acting on the USC system, $\mathcal{D}[\mathcal{O}]\rho=1/2(2\mathcal{O}\rho \mathcal{O}^\dagger-\rho \mathcal{O}^\dagger \mathcal{O}-\mathcal{O}^\dagger \mathcal{O}\rho)$ is the dissipator, and $\tau^x_m$, $\tau^z_m$ correspond to the noise channels of the $m$-th transmon, which dissipates with ratio $\gamma_m$, and loses coherence at rate $\gamma_{\phi_m}$, respectively. Moreover, $\Gamma^{jk}_{\uparrow{\rm{eff}}}$, and $\Gamma^{jk}_{\downarrow{\rm{eff}}}$ correspond to the dressed heating and decay rates in the mediator, given by
\begin{align}\nonumber
\Gamma^{jk}_{\downarrow{\rm{eff}}}&= \bigg[\Gamma^{jk}_\kappa+\sum_{\ell=1}^2\bigg(\Gamma^{jk}_{\gamma,\ell}+\Gamma_{\phi,\ell}^{jk}\bigg)\bigg](\bar{n}(\Delta_{kj},T)+1),\\
\Gamma^{jk}_{\uparrow{\rm{eff}}}&= \bigg[\Gamma^{jk}_\kappa+\sum_{\ell=1}^2\bigg(\Gamma^{jk}_{\gamma,\ell}+\Gamma_{\phi,\ell}^{jk}\bigg)\bigg]\bar{n}(\Delta_{kj},T),~~~~~~
\end{align}
where $\bar{n}(\Delta_{kj},T)=1/(\exp(\hbar\Delta_{kj}/k_BT)-1)$ is the number of thermal photons of energy $\hbar\Delta_{kj}$ at temperature $T$~\cite{Settineri2018}, with $k_{B}$  the Boltzmann constant.  $\Gamma^{jk}_\kappa$ is the relaxation rate provided by the resonator, and $\Gamma^{jk}_{\gamma,\ell}$ is the relaxation rate of the $\ell$-th charge qubit due to the coupling with its thermal bath. Finally, $\Gamma_{\phi,\ell}^{jk}$ stands for the effective relaxation rate due to the fact that $\sigma^{z}$ is not diagonal in the dressed basis. These rates are defined as   
\begin{subequations}
\begin{align}
\Gamma^{jk}_\kappa&=\frac{\kappa\Delta_{kj}}{{\omega}_{r}}|\langle \psi_{j}|{a}+{a}^\dagger|\psi_{k}\rangle|^2,\\
\label{re_LC}
\Gamma^{jk}_{\gamma,\ell}&=\frac{\gamma\Delta_{kj}}{\omega_{q,\ell}}|\langle \psi_{j}|\sigma^x_\ell|\psi_{k}\rangle|^2,\\
\Gamma_{\phi,\ell}^{jk}&=\frac{\gamma^{\textrm{CPB}}_\varphi \Delta_{kj}}{\omega_{q,\ell}}|\langle \psi_{j}|\sigma^z_\ell|\psi_{k}\rangle|^2,
\label{re_cpb}
\end{align}
\end{subequations}

where $\kappa$, $\gamma$, and $\gamma_{\phi}^{{\textrm{CPB}}}$ are the bare decay rates corresponding to cavity leakage, energy loss on the two-level systems, and the depolarizing noise on the qubit, respectively. In this case, we have assumed that the energy loss on the mediator subsystem is provided by an ohmic spectral density bath, which is an archetypical noise source on these systems~\cite{Settineri2018}. Fig.~\ref{matrix_population}{(a)} show the ratio between the dressed decay rates \ref{re_LC} and \ref{re_cpb} with its bare ones for the three different coupling strength values. From the figure, we appreciate that depending on the coupling strength, different states take part in the  dissipative dynamics. On the other hand, the dressed dephasing rates are 
\begin{align}
\Gamma_{\phi,\ell}^{jj}=\frac{\gamma^{\textrm{CPB}}_\varphi}{2\omega_{q,\ell}}\omega_T|\langle \psi_{j}|\sigma^z_\ell|\psi_{j}\rangle|^2,\label{eq:dressed_deph}
\end{align}
where the thermal frequency $\omega_T=k_BT/\hbar$.  For the frequencies and temperatures considered in Fig.~\ref{matrix_population}, such rates Eq.~(\ref{eq:dressed_deph}) are very small. This result connects to the possibility of enhancing the coherence time of logical quantum information encoded in a USC unit studied in~\cite{Nataf2011}, where the states of the CPB are highly entangled with the ones of photons in the oscillator, which are robust concerning a general class of ‘‘anisotropic’’ environment. To obtain a master equation, we have considered a regime where the charge noise in the CPBs is the dominant contribution to the loss of quantum coherence~\cite{Astafiev2004}, where, in the long time limit, can also be approximated by Gaussian noise~\cite{Cottet2002,Koch2007,Zaretskey2013,Paladino2014}.\par 
In order to perform the numerical simulations, we prepare the system in the state $\rho_{1}=\ket{1}\bra{1}\otimes\rho^{{\rm{Th}}}\otimes\ket{0}\bra{0}$, where the mediator is initialized in the thermal state $\rho^{\rm{Th}}=\exp (-\beta H_{\rm{coupler}})/\mathcal{Z}_{{{\rm{coupler}}}}$, where $\beta=1/k_{B}{T}$, and $\mathcal{Z}={\rm{Tr}}[\exp(-\beta H)]$ is the partition function and compare with the state $\rho_{2}=\ket{0}\bra{0}\otimes\rho^{\rm{Th}}\otimes\ket{1}\bra{1}$ under loss mechanism. The calculations have been performed with the experimentally motivated parameters $\omega_{r}/2\pi=8.13\textrm{\,GHz}$~\cite{FornDiaz2016}, $\gamma_m/2\pi=0.48$\,MHz, $\gamma_{\phi_m}/2\pi=0.15$\,MHz, $\kappa/2\pi=0.10$\,MHz~\cite{Saira2014}, $\gamma/2\pi=0.0083$\,MHz~\cite{Zaretskey2013}, $\gamma_\phi^{\textrm{CPB}}/2\pi=2.00$\,MHz~\cite{Metcalfe2007}, and the system is at $T=50~{\rm{mK}}$, i.e., the thermal frequency $\omega_T/2\pi=1.042$\,GHz.

In the QST protocol, we may expect that a significantly greater coupling strength on the mediator leads to better performance on the state transfer. Nevertheless, this is not the situation we observe in Fig.~\ref{matrix_population}{(b)}; in the pUSC regime, the QST protocol is successful with transfer fidelity equal to $\mathcal{F}=0.983$ at transfer time $t=424.8~\rm{ns}$. Let us compare this result with the mediator operating beyond this regime. It is possible to appreciate better transfer fidelity $\mathcal{F}=0.986$ in a shorter gate time ($t=49.3~\rm{ns}$) that protects the qubit state against the interaction with the reservoir. This case compares well with the previous proposal based on Flux-qubits~\cite{Cardenas2017}. Thus, we may expect that transfer fidelity and gate time in the DSC regime should decrease even more. However, the QST protocol got a worse result than the previous two cases in state-transfer fidelity $\mathcal{F}=0.692$, but with transfer time $t=498.5~\rm{ns}$ similar to the one obtained in the pUSC regime. 

The answer to this counter-intuitive behavior lies in the level structure of the mediator. In Fig.~\ref{spectrum}(a), we observe that at $\tilde{g}_{\ell}/\omega_{r}=0.5$, the relative anharmonicity of the energy states is greater than in the pUSC at $\tilde{g}_{\ell}/\omega_{r}=0.3$. In the later regime, we observe a shrunk in the energy levels of the system, leading to an enhancement of the number of states of the mediator involved in the dynamics, yielding a reasonable performance of the QST protocol. This situation is more remarkable in the DSC regime ($\tilde{g}_{\ell}/\omega_{r}>1$), where the energy spectrum tends towards that of a displaced oscillator where every energy manifold is double degenerate. Such degeneration of the energy spectrum yields that, for a particular resonance condition, there are more mediator states participating in the effective interaction.

Let us stress again that this protocol strongly depends on the level structure of the mediator and its coupling operators to the adjacent information-exchanging partners, two combined properties native to the USC regime~\cite{Nataf2011}. In this case, we have used the particular light-mode forbidden transitions of the USC-unit, but any other analogous scheme could be potentially implemented. Using the same proposed circuit, but working in the strong coupling regime (where the RWA is applicable), will not produce any similar results.

\section{Conclusions}
\label{sec5}
We have performed a thorough numerical study of the feasibility to construct charge qubits coupled to LC harmonic oscillators in the USC-DSC regime, generalizing previous seminal works~\cite{Nataf2010,Jaako2016,ManucharyanJPA2017}. In doing so, we have optimized the macroscopic circuital parameters for enhancing a series-parallel coupling configuration while maintaining a resonance condition. 

Later, we have used this compound CPB-based USC unit as an incoherent (thermally populated) mediator to perform the quantum state transfer protocol, improving a previous proposal with flux qubits. The key ingredients for obtaining comparable transfer times without requiring a highly coherent and controllable mediator are the combination of an anharmonic energy spectrum and particular selection rules for the coupling operators, both features being inherent of the USC regime. Our theoretical results encourage further experimental exploration of the physics of the light-matter interaction with large couplings and multi-mode resonators and possible applications for quantum simulation and quantum information processing.

\section*{Acknowledgements}
We thank J. Braum\"uller, P. Forn-D\'iaz and I. L. Egusquiza for enlightening discussions and comments on the manuscript. The authors acknowledge support from Spanish MCIU/AEI/FEDER (PGC2018-095113-B-I00), Basque Government IT986-16, projects QMiCS (820505) and OpenSuperQ (820363) of EU Flagship on Quantum Technologies, EU FET Open Grants Quromorphic and EPIQUS, Shanghai STCSM (Grant No. 2019SHZDZX01-ZX04). F. A. C. L thanks to the German Ministry for Education and Research, under QSolid, Grant no. 13N16149. A. P.-R. thanks Basque Government Ph.D. grant PRE-2016-1-0284. F. A. C. L. acknowledge support from
the German Ministry for Education and Research, under
QSolid, Grant no. 13N16149. We thank the referees for a thorough criticism of the manuscript.

\appendix
\section{One CPB coupled to an LC oscillator}\label{appendixA}
For completeness, we derive the Hamiltonian of the one-qubit system depicted in Fig.~\ref{model}(a) by first computing the Lagrangian with node-flux variables 
\begin{align}
L_{1}  = & \frac{C_p}{2}\dot{\Phi}_1^2 + \frac{C_r}{2}\dot{\Phi}_r^2+\frac{C_c}{2}(\dot{\Phi}_r+\dot{\Phi}_1)^2+ \frac{C_J}{2}\dot{\Phi}_J^2 \label{eqA1}\\ &+\frac{C_g}{2}(\dot{\Phi}_J+\dot{\Phi}_r+\dot{\Phi}_1-V_g)^2-\frac{\Phi_r^2}{2L_r}+E_J\cos{\big(\varphi_J\big)},\nonumber
\end{align}
where $\Phi_p$, $\Phi_{r}$, and $\Phi_{J}$ are the fluxes describing the capacitance $C_p$, the LC oscillator, and the CPB, respectively. The flux velocity $\dot{\Phi}_{1}$ corresponds to a passive node (no potential energy associated with the coordinate $\Phi_p$) which can be eliminated noticing that the Euler-Lagrange equation reads $\partial_t( \partial L_{1}/\partial[\dot{\Phi}_{1}])=0$. This relation allows us to write $\dot{\Phi}_{1}$ in terms of the remaining fluxes 
\begin{align}
\dot{\Phi}_{1} = \frac{-C_{g}\dot{\Phi}_{J}-(C_{c}+C_{g})\dot{\Phi}_{r}}{\tilde{C}}+V_{1}(0),
\end{align}
where we have defined $\tilde{C}=C_c+C_g+C_p$, and $V_{1}(0)=\dot{\Phi}_{1}(0)$ is an initial voltage bias (integration constant). By replacing this expression in Eq. (\ref{eqA1}), we can obtain the Lagrangian only in terms of $\Phi_{r}$ and $\Phi_{J}$
\begin{align}
L_{1}=\frac{1}{2}\dot{\vec{\Phi}}^{T}\mathbfcal{C}\dot{\vec{\Phi}}-V_{g}\dot{\vec{\Phi}}^{T}\mathbfcal{C}_{v}-\frac{\Phi_r^2}{2L_r}+E_J\cos{\big(\varphi_J\big)},
\label{EqA4}
\end{align}
where we have defined the coordinates' vector as $ \vec{\Phi}=(\Phi_J, \Phi_{r})^T$, and we discarded the terms proportional to $V_{g}^{2}$ and $V_1(0)^2$, which do not contribute to the dynamics. $\mathbfcal{C}$ and $\mathbfcal{C}_v$ are the capacitance matrix and the gate capacitance vector, respectively, defined as 
\begin{align}
&\mathbfcal{C} =\frac{1}{\tilde{C}}\nonumber
\begin{pmatrix}
\tilde{C}C_J + {C_{g}(C_{c}+C_{p})}&{C_{g}C_{p}}\\
{C_{g}C_{p}}&\tilde{C}C_{r}+ {C_{p}(C_{c}+C_{g})}
\end{pmatrix},\\
&\mathbfcal{C}_v =\frac{1}{\tilde{C}}
\begin{pmatrix}
C_{g}(C_{c}+C_{p})\\
C_{g}C_{p} \\
\end{pmatrix}.\nonumber
\end{align}
We compute the canonical conjugate momenta to the flux variables  $\vec{Q}=\partial L/\partial[\dot{\vec{\Phi}}^{T}]=\mathbfcal{C}\dot{\vec{\Phi}}-V_{g}\mathbfcal{C}_v$, and apply the Legendre transformation to obtain the Hamiltonian
\begin{align}\nonumber
H_{1} =& \frac{1}{2}(\vec{Q}+V_g\mathbfcal{C}_v)^T\mathbfcal{C}^{-1}(\vec{Q}+V_g\mathbfcal{C}_v)-E_J\cos{\varphi_J}\\
= & \frac{(Q_J-2 e n_g)^2}{2 \tilde{C}_J}-E_J\cos{\big(\varphi_J\big)}+\frac{Q_r^2}{2\tilde{C}_r} +\frac{\Phi_r^2}{2L_r}\nonumber\\
&+\frac{Q_r(Q_J-2e n_g)}{C_{Jr}},
\label{EqA7}
\end{align}
where we have defined the capacitances
\begin{align}
\tilde{C}_J=\frac{1}{[\mathbfcal{C}^{-1}]_{1,1}}, \tilde{C}_r=\frac{1}{[\mathbfcal{C}^{-1}]_{2,2}},{C}_{Jr}=\frac{1}{\big|[\mathbfcal{C}^{-1}]_{1,2}\big|}.
\label{eq:def_CJ_Cr_CJr_tildes}
\end{align}
Please notice that the charge biased has been removed from the oscillator degree of freedom simply by shifting the charge coordinates $Q_r+a\rightarrow Q_r$, $\forall a\in \mathbbm{R}$ which does not affect the commutation relations. We have defined $n_g=V_g C_g(C_c + C_p)/(2 e\tilde{C})$. In this article, we always assume that the CPBs work on their sweet spot, i.e., $n_g=0.5$. It is worth noting that the charge bias can also be removed from the Josephson-junction sector of the Hamiltonian, at the expense of modifying the periodic boundary conditions.

We canonically quantize the circuit Hamiltonian by promoting the conjugated variables to quantum operators, $\hat{\Phi}_{r} = i\sqrt{\frac{\hbar {\omega}_{r} L_r}{2}}(a-a^\dagger)$, $\hat{Q}_{r} = \sqrt{\frac{\hbar {\omega}_{r}\tilde{C}_r}{2}}(a+a^\dagger)$, $\hat{Q}_J=2e \hat{n}_J$, and $e^{i\hat{\varphi}_J}=\sum_{n_J}\ket{n_{J}+1}\bra{n_{J}}$, where $\omega_r=1/\sqrt{\tilde{C}_rL_r}$ is the frequency of the resonator, and $a$ ($a^{\dag}$) are the annihilation (creation) operators of the quantum harmonic oscillator satisfying the commutation relation $[a,a^{\dag}]=1$, such that $[\hat{\Phi}_{r},\hat{Q}_{r}]=i\hbar$.
The Cooper-pair number $\hat{n}_J$ and phase $\hat{\varphi}_{J}$ operators satisfy the commutation relation $[e^{i\hat{\varphi}_{J}},\hat{n}_{J}]=e^{i\hat{\varphi}_{J}}$. After the above transformations, we arrive at the Hamiltonian in Eq. (\ref{Eq04}) of the main text
\begin{align}\nonumber
H_1 = & 4E_C (\hat{n}_J-n_g)^2 - E_J \cos(\hat{\varphi}_J)+ \hbar{\omega}_{r} a^\dagger a \\
&+ \hbar g(\hat{n}_J-n_g)(a + a^\dagger),\label{eq:H_1_appendix_final}
\end{align}
where $E_C=e^2/2\tilde{C}_{J_1}$ is the charge energy, and the coupling strength between the CPB and the resonator reads
\begin{align}
 g &= {\omega}_{r}\gamma \sqrt{\frac{\tilde{Z}_{r}}{2R_Q}},\nonumber
\end{align}
where $\gamma=\tilde{C}_r/C_{Jr}$, $\tilde{Z}_{r} = \sqrt{L_r/\tilde{C}_r}$ is the impedance of the effective resonator mode and $R_Q = \hbar/(2e)^2$ is the reduced quantum resistance. It is worth noticing that the Hamiltonian description of the circuit here studied is analogous, but more general, to that obtained in Jaako {\it et al.}~\cite{Jaako2016} for the single CPB case.

\section{Two-oscillator model}
\label{appendix_two_oscillator_model}
In this section, we show that there exists a fundamental limit for the coupling strength between two harmonic systems. For doing so, we consider the circuit depicted in Fig.~\ref{model}(a), and we change the Josephson junction of the CPB by a linear inductor $L_J$, which allows us to gauge out both the gate voltage $V_{g}$ completely. The Hamiltonian for this system reads
\begin{align}
H_{o}=\frac{Q_J^2}{2\tilde{C}_J}+ \frac{\Phi_J^2}{2L_J}+\frac{Q_r^2}{2\tilde{C}_r} + \frac{\Phi_r^2}{2L_r}-\frac{Q_rQ_J}{C_{Jr}},\nonumber
\end{align}
which has a similar expression to Hamiltonial (\ref{eq:H_1_appendix_final}) except that we replace the Josephson energy term with the potential energy ${\Phi_J^2}/{2L_J}$, and neglect the terms proportional to $V_g^2$. The dressed capacitances $\tilde{C}_J$, $\tilde{C}_r$, and $C_{Jr}$ are identical to those defined in Eq.~(\ref{eq:def_CJ_Cr_CJr_tildes}). Next, we quantize the Hamiltonian $H_{o}$ by promoting the classical conjugated variables to quantum operators, $\hat{Q}_J=\sqrt{\hbar \omega_q \tilde{C}_J/2}(b+b^\dagger)$, $\hat{\Phi}_{J} = i\sqrt{\hbar {\omega}_{q} L_J/2}(b-b^\dagger)$, which satisfy the commutation relation $[\hat{\Phi}_{J},\hat{Q}_{J}]=i\hbar$, and $\omega_q=1/\sqrt{\tilde{C}_JL_J}$ is the frequency of the oscillator, where $b$ ($b^{\dag}$) is the corresponding annihilation (creation) operator satisfying the bosonic commutation relation $[b,b^{\dag}]=1$. For the resonator, we use the procedure as in the previous Appendix. The quantized Hamiltonian reads 
\begin{align}
\hat{H}_{o}=\hbar \omega_q b^\dagger b+\hbar \omega_r a^\dagger a-\hbar g_{o}(a+a^\dagger)(b+b^\dagger),\nonumber
\end{align}
where the coupling strength between the harmonic system is defined as 
\begin{align}
 g_{o}&=\frac{\gamma_{o}}{2}\sqrt{ \omega_r \omega_q },\nonumber
\end{align}
with the dimensionless ratio 
\begin{align}
\gamma_{o}=\sqrt{ \frac{\tilde{C}_J\tilde{C}_r}{C^2_{J_r}}}=\sqrt{\frac{1}{(1+\frac{C_gC_c+C_J\tilde{C}}{C_gC_p})(1+\frac{C_pC_c+C_r\tilde{C}}{C_gC_p})}}\nonumber
\end{align}
written in terms of the bare system capacitances. Notice that in the limit $C_g, C_p\gg C_c, C_J$, $\gamma_{o}\rightarrow 1$. For generic values of the capacitors, we see that ${2g_{o}}/{\sqrt{\omega_q\omega_r}}\leq 1$, i.e., there exists a fundamental limit for the value of the coupling strength between the subsystems.

\section{Two CPBs coupled to an LC oscillator}{\label{appendix_series_CPB_Ham}}
We derive here the Hamiltonian of the two-qubit system, as shown in Fig.~\ref{model}(b) considering identical grounded capacitances \ie $C_{c0} =C_{c1}=C_c$. We first calculate the Lagrangian of the circuit following the same procedure as in Appendix~\ref{appendixA}, i.e., writing the passive node $\Phi_0$ in terms of $\Phi_{J1}$, $\Phi_{J2}$, and $\Phi_{r}$ to obtain
\begin{align}\nonumber
L_{2} = & \frac{1}{2}\dot{\vec{\Phi}}^{T}\mathbfcal{C}\dot{\vec{\Phi}}-\tilde{V}_{g1}\dot{\vec{\Phi}}^{T}\mathbfcal{C}_{v1}-\tilde{V}_{g2}\dot{\vec{\Phi}}^{T}\mathbfcal{C}_{v2}-\frac{\Phi_r^2}{2L_r}\\
&+E_{J1}\cos{\big(\varphi_{J1}\big)}+E_{J2}\cos{\big(\varphi_{J2}\big)},
\end{align}
again neglecting the constant term, where $\vec{\Phi}=(\Phi_{J1}, \Phi_{J2}, \Phi_{r})^T$ is the flux vector. The capacitance matrix and the gate capacitance vectors are now
\begin{widetext}
\begin{align}\nonumber
&\mathbfcal{C} =\frac{1}{\tilde{C}}
\begin{pmatrix}
\tilde{C}{C_{J1}}+{C_{g1}(2C_c+C_{g2})}&{C_{g1}C_{g2}}&{C_{g1}(C_{g2}+C_c)}\\
C_{g1}C_{g2}&\tilde{C}C_{J2}+C_{g2}(2C_c+C_{g1})&C_{g2}(C_c+C_{g1})\\
C_{g1}(C_{g2}+C_c)&C_{g2}(C_c+C_{g1})&\tilde{C}C_r+(C_c+C_{g1})(C_c+C_{g2})
\end{pmatrix},\\
&\mathbfcal{C}_{v1}=-\frac{1}{\tilde{C}}
\begin{pmatrix}
{C_{g1}(2C_c+C_{g2})}\\
{C_{g1}C_{g2}}\\
{C_{g1}(C_c+C_{g2})}\\
\end{pmatrix},\quad
\mathbfcal{C}_{v2} =
-\frac{1}{\tilde{C}}\begin{pmatrix}
{C_{g1}C_{g2}}\\
{C_{g2}(2C_c+C_{g1})} \\
{C_{g2}(C_c+C_{g1})}\\
\end{pmatrix},
\label{EqB4}
\end{align}
\end{widetext}
where we have now $\tilde{C}=2C_c + C_{g1}+C_{g2}$. We again apply the Legendre transformation, and we eliminate the interaction between the gate voltages and the resonator  analogously to Appendix~\ref{appendixA} obtaining
\begin{align}\nonumber
H_2 = & \sum^2_{\ell=1}\bigg[\frac{\mathcal{Q}_{J\ell}^2}{2\tilde{C}_{J\ell}}-\frac{Q_r\mathcal{Q}_{J\ell}}{C_{2\ell}}-E_{J\ell}\cos{(\varphi_{J\ell})}\bigg]\\
&+\frac{Q_r^2}{2\tilde{C}_r}+\frac{\Phi_r^2}{2L_r}-\frac{\mathcal{Q}_{J1}\mathcal{Q}_{J2}}{C_{12}},
\label{EqB5}
\end{align}
with $\mathcal{Q}_{J\ell}=\big(Q_{J\ell}-2en_{g\ell}\big)$, where $n_{g\ell}$ is the effective gate-charge number of the $\ell$-th CPB, which is tunable through the gate voltages and again assumed to be at the sweet spot, i.e., $n_{g\ell}=0.5$. The effective Josephson, resonator, and coupling capacitances are defined as 
\begin{align}\nonumber
&\tilde{C}_{J\ell}=\frac{1}{[\mathbfcal{C}^{-1}]_{\ell,\ell}}, \tilde{C}_r=\frac{1}{[\mathbfcal{C}^{-1}]_{3,3}},\\
&{C}_{\ell r}=\frac{1}{\big|[\mathbfcal{C}^{-1}]_{\ell,3}\big|}, {C}_{12}=\frac{1}{\big|[\mathbfcal{C}^{-1}]_{1,2}\big|},\nonumber
\end{align}
respectively, where $[\mathbfcal{C}^{-1}]_{j,k}$ is the $jk$-th element of the inverse capacitance matrix. After the quantization procedure and a unitary transformation we arrive to Hamiltonian
\begin{align}\nonumber
{H}_2 = & \sum_{\ell=1}^2 \hat{H}^{\rm{cpb}}_{\ell}+ \hbar \omega_{r}{a}^\dagger {a}+\hbar\sum^2_{\ell=1}  g_\ell(\hat{n}_{J\ell}-n_{g\ell})({a}+{a}^\dagger)\\
&+\hbar g_{12}(\hat{n}_{J1}-n_{g1})(\hat{n}_{J2}-n_{g2}),\nonumber
\end{align}
where $\hat{H}^{\rm{cpb}}_{\ell}=4E_{C\ell} (\hat{n}_{J\ell}-n_{g\ell})^2 - E_{J\ell}\cos(\hat{\varphi}_{J\ell})$ corresponds to the Hamiltonian of the $\ell$-th CPB, with the charge energy $E_{C\ell}=e^2/2\tilde{C}_{J\ell}$ and the coupling strengths
\begin{align}
 g_{\ell}&=\omega_r\gamma_\ell\sqrt{\frac{\tilde{Z}_r}{2R_Q}},\quad
 g_{12}= -\frac{1}{R_Q C_{12}},\nonumber
\end{align}
which correspond to the interaction between the $\ell$th CPB and the LC resonator, and the two CPBs, respectively. Moreover, the ratio $\gamma_\ell=\tilde{C}_r/C_{\ell r}$.

\section{Derivation of the Master Equation}{\label{appendixD}}
Here we derive the master equation (ME) to describe the interaction between the transmons-USC unit and the environment, which is modeled as an infinite set of quantum harmonic oscillators~\cite{Beaudoin2011}.  We include the effects of energy loss/gain in the  transmons, CPBs, and LC oscillator, and the depolarizing channel for each CPB independently. In ultrastrongly coupled systems, the standard (perturbative) master equation description leads to unphysical results~\cite{Settineri2018}. To avoid this, we express the USC operators in the dressed basis of the whole system. 
\subsection{Energy relaxation}
Let us begin by obtaining a ME for the energy relaxation process, following the procedure of Beaudoin et al.~\cite{Beaudoin2011}, where we assume that the USC system is weakly coupled to the environment through the $X$ quadrature of the LC oscillator and $\sigma_{\ell}^x$ for the $\ell$th CPB. We describe this situation with the following Hamiltonian ($\hbar=1$)
\begin{align}\label{EqC1}
H_{SE} &= H_\textrm{USC} + H_{B} + H_{SB}\\\nonumber
&= \sum_{j}\varepsilon_j\sigma_{jj} + \sum_l\nu_lb_l^\dagger b_l + \sum_l(c+c^\dagger)(h_lb_l+h_l^* b_l^\dagger) .
\end{align}
Here, the first term corresponds to the USC Hamiltonian expressed it in its diagonal basis, with the projector $\sigma_{nm}=\ket{\psi_n}\bra{\psi_m}$, the second term to the Hamiltonian describing the infinite quantum harmonic oscillator with frequency $\nu_{l}$, and $b_l(b_l^\dagger)$ is the ladder operator for the $l$th bath mode, and the last terms to the system-environment interaction. Depending on whether we are calculating relaxation rates for the resonator or the CPBs, the operator $(c+c^{\dag})$ will correspond to the field quadrature $X$ for the resonator, or to the $x$-Pauli matrix for the $\ell$-th CPB, expressed in the dressed basis as $c+c^\dagger=\sum_{jk}C_{jk}\sigma_{jk}$, with the real and symmetric matrix element $C_{jk}=C^*_{jk}=\bra{\psi_j}c+c^\dagger\ket{\psi_k}$. In the interaction picture, it takes the form
\begin{align}
\tilde{H}_{SB}(t)=\sum_{jkl}C_{jk}\sigma_{jk}(h_lb_le^{-i\nu_lt}+h_l^*b_l^\dagger e^{i\nu_lt})e^{i\Delta_{jk}t},
\label{EqC2}
\end{align}
where $\Delta_{jk}=\varepsilon_j-\varepsilon_k$ is the energy transition between the $j$th and $k$th energy level of the USC system. Now, we split the interaction Hamiltonian in Eq.~(\ref{EqC2}) into two parts
\begin{align}\nonumber
\tilde{H}_{SB}(t) = & \sum_{j,l}C_{jj}\sigma_{jj}(h_lb_le^{-i\nu_lt}+h_l^*b_l^\dagger e^{i\nu_lt})\\
&+\sum_l\sum_{j,k>j}C_{jk}\sigma_{jk}\bigg[\big(h_lb_le^{-i(\nu_l+\Delta_{kj})t} \nonumber \\ & +h_l^*b_l^\dagger e^{i(\nu_l-\Delta_{kj})t}\big)\nonumber\\
&+\big(h_lb_le^{-i(\nu_l-\Delta_{kj})t}+h_l^*b_l^\dagger e^{i(\nu_l+\Delta_{kj})t}\big)\bigg].
\label{EqC3}
\end{align}
We can simplify the interaction Hamiltonian defining the operators
\begin{align}
\tilde{S}_1=\sum_{j,k>j}S_1e^{-i\Delta_{kj}t}, \tilde{S}_2=\sum_{j,k>j}S_2e^{i\Delta_{kj}t},
\label{EqC4}
\end{align}
where $S_1=C_{jk}\sigma_{jk}$, $S_2=C_{jk}\sigma_{kj}$, and  ${S}_3=\sum_{j}C_{jj}\sigma_{jj}$, and similarly for bath operators
\begin{align}
\tilde{B}_1(t)=\sum_l h_l^* b_l^\dagger e^{i\nu_l t},\quad \tilde{B}_2(t)=\sum_l h_l b_le^{-i\nu_l t}.
\label{EqC5}
\end{align}

By replacing Eq.~(\ref{EqC4}), and Eq.~(\ref{EqC5}) in Eq.~(\ref{EqC3}), we obtain 
\begin{align}
\tilde{H}_{SB}(t)&=\big(\tilde{S}_1+\tilde{S}_2+S_3\big)\big(\tilde{B}_1(t)+\tilde{B}_2(t)\big).\nonumber
\end{align}
In the interaction picture, the dynamics of the system-bath Hamiltonian is described by the von Neumann equation
\begin{align}
\frac{d\tilde{\rho}_{SB}(t)}{dt}=-i[\tilde{H}_{SB}(t),\tilde{\rho}_{SB}(t)].\nonumber
\end{align}
The formal solution of this equation reads 
\begin{align}
\tilde{\rho}_{SB}(t+\Delta t)= \tilde{\rho}_{SB}(t)-i\int_t^{t+\Delta t} dt_1[\tilde{H}_{SB}(t_1),\tilde{\rho}_{SB}(t_1)].\nonumber
\end{align}
We replace this solution on the von Neumann equation, and we trace over the degrees of freedom of the environment for approximately describing the dynamics of the subsystem, obtaining
\begin{align}\label{EqD6}
{\Delta\tilde{\rho}(t)} = & \frac{1}{i}\int_{t}^{t+\Delta t}dt_1 \textrm{Tr}_B[\tilde{H}_{SB}(t_1),\tilde{\rho}(t)\otimes \bar{\rho}_B] \\ \nonumber & -\int_t^{t+\Delta t} dt_1\int^{t_1}_{t}dt_2 \textrm{Tr}_B\Big[ \tilde{H}_{SB}(t_1), \\ \nonumber & [\tilde{H}_{SB}(t_2),\tilde{\rho}(t)\otimes \bar{\rho}_B]\Big],
\end{align}
where we have retained the terms in the interaction Hamiltonian up to second order. The standard way to proceed here, is to introduce two approximations to simplify this equation. Firstly, we assume that the state of the system-environment is separable $\forall\, t$, also known as the Born approximation~\cite{Breuer2002}, i.e.,  $\tilde{\rho}_{SB}(t)=\tilde{\rho}(t)\otimes\bar{\rho}_B$, where $\tilde{\rho}(t)$ and $\bar{\rho}_B$ are the density matrices of the USC and bath subsystems, respectively. Secondly, we apply the Markov approximation, which presumes different time-scales for the system and environment dynamics. 

Let us perform the change of variable $\tau=t_1-t_2$, leading to $\int_t^{t+\Delta t} dt_1\int^{t_1}_{t}dt_2=\int_0^{\Delta t} d\tau\int^{t+\Delta t}_{t+\tau}dt_1$, where $\Delta t$, and $\tau$ are the time-scales describing the dynamics of the system and the environment, respectively. The Markov approximation states that $\tau\ll\Delta t$ meaning that the correlation function's time is shorter than the system ones. This allow us to change the limit of integration of both integrals as follows~\cite{Kryszewski2008} 
\begin{widetext}
\begin{align}\nonumber
\frac{\Delta\tilde{\rho}(t)}{\Delta t} = & \sum_{\Delta_{kj},\Delta_{k^\prime j^\prime}}  \frac{1}{\Delta t}\Bigg\{ \int_0^{\infty}d\tau e^{i\Delta_{kj}\tau}  \bar{G}_{11}(\tau)  \int_{t}^{t+\Delta t}dt_1 \ e^{i(\Delta^\prime_{kj}-\Delta_{kj})t_1} \bigg[S_1\tilde{\rho}(t)S_1^{\prime^\dagger} - S_1^{\prime^\dagger}S_1 \tilde{\rho}(t)\bigg] \\ \nonumber
&+\int_0^{\infty}d\tau e^{-i\Delta_{kj}\tau}  \bar{G}_{22}(\tau)  \int_{t}^{t+\Delta t}dt_1\ e^{i(\Delta_{kj}-\Delta^\prime_{kj})t_1} \bigg[S_2\tilde{\rho}(t)S_2^{\prime^\dagger}-S_2^{\prime^\dagger} S_2 \tilde{\rho}(t)\bigg]\Bigg\}
+\rm{{H.c.}},
\end{align}
\end{widetext}
where we have performed the secular approximation neglecting the fast oscillating terms proportional to $\textrm{exp}\{{\pm i\Delta_{k^\prime j^\prime}t_1}\}$, $\textrm{exp}\{{\pm i\Delta_{kj}t_2}\}$, $\textrm{exp}\{{\pm i(\Delta_{k^\prime j^\prime}t_1}+\Delta_{kj}t_2)\}$, $\textrm{exp}\{\pm i\nu_l \tau\}$, and $\textrm{exp}\{\pm i(\nu_l+\Delta_{kj})\tau\}$. In addition, we have defined $S^\prime_1=C_{j^\prime k^\prime}\sigma_{j^\prime k^\prime}$, $S^\prime_2=C_{j^\prime k^\prime }\sigma_{k^\prime j^\prime}$. Notice that for an ohmic bath we can neglect the contributions coming from $S_3$~\cite{Settineri2018}. Finally, the correlations functions of the bath $\bar{G}_{\alpha\beta}(\tau)=\textrm{Tr}[\tilde{B}^\dagger_{\alpha}(\tau){B}_{\beta}\bar{\rho}_{B} ]$ reads
\begin{align}\label{cor}
\bar{G}_{11}(\tau)&=\sum_l e^{-i\nu_l \tau} |h_l|^2\bigg[\bar{n}(\nu_l,T)+1\bigg],\\\nonumber
\bar{G}_{22}(\tau)&=\sum_l e^{i\nu_l \tau} |h_l|^2\bar{n}(\nu_l,T),
\end{align}
where $\bar{n}(\nu_l,T)=[e^{{\nu_l}/k_BT}-1]^{-1}$ is the average of thermal photons of the $l$th mode~\cite{Settineri2018}. In the Schr$\ddot{\textrm{o}}$dinger picture, we see that the only terms contributing correspond to $k=k^\prime$, and $j=j^\prime$~\cite{Kryszewski2008} (second secular approximation) leading to
\begin{align}\nonumber
\frac{d{\rho}(t)}{dt} = & -{i}[H_\textrm{USC}+H^1_{LS},{\rho}(t) ]\\\nonumber
&+ \sum_{\Delta_{kj}}  W_{11}(\Delta_{kj}) \bigg[S_1{\rho}(t) S_1^{\dagger} - S_1^{\dagger}S_1 {\rho}(t) \bigg]\\\nonumber
&+\sum_{\Delta_{kj}}W_{22}(-\Delta_{kj}) \bigg[S_2 {\rho}(t)S_2^{\dagger}-S_2^{\dagger} S_2 {\rho}(t)\bigg]
+\rm{{H.c}}.,
\end{align}
with $W_{11}(\Delta_{kj})=\int_0^{\infty}d\tau e^{i\Delta_{kj}\tau} \bar{G}_{11}(\tau)$, and $W_{22}(-\Delta_{kj})=\int_0^{\infty}d\tau e^{-i\Delta_{kj}\tau} \bar{G}_{22}(\tau)$ corresponding to the Fourier transform of the correlation function of the bath. Now we write the ME in its standard form as follows
\begin{align}
&\frac{d}{dt}{\rho}(t)=-{i}[{H}_\textrm{USC}+{H}^1_{LS},{\rho}(t) ]+ \sum_{\Delta_{kj}}  \mathcal{J}(\Delta_{kj})\\
&\bigg[[\bar{n}(\Delta_{kj},T)+1] \mathcal{D}[{S}_1]\rho(t)+\bar{n}(\Delta_{kj},T) \mathcal{D}[S_2]\rho(t)\bigg],\nonumber
\label{EqC13}
\end{align}
where $\mathcal{D}[\mathcal{O}]\rho(t)=1/2[2\mathcal{O}\rho(t) \mathcal{O}^\dagger -[\mathcal{O}^\dagger \mathcal{O},\rho(t)]_{+}]$ is the Lindbladian, $\mathcal{J}(\omega)=\sum_l 2\pi|h_l|^2\delta(\omega-\nu_l)$ is the spectral density of the bath, and $\mathcal{H}^1_{LS}$ is the Lamb-shift Hamiltonian
\begin{align}
{H}^1_{LS}=\sum_{\Delta_{kj}}\delta_{11}(\Delta_{kj})S^\dagger_1 S_1+\delta_{22}(-\Delta_{kj})S_2^{\dagger} S_2,
\end{align}
with $\delta_{\alpha\alpha}({\omega})=-i[W_{\alpha\alpha}(\omega)-W^*_{\alpha\alpha}(\omega)]/2$. We recall that in the weak-coupling limit, and upon the use of a system-environment coupling model with a natural but far-enough cutoff~\cite{ParraRodriguez2018}, the Lamb-shift terms commute with the system Hamiltonian~\cite{Breuer2002} just modifying the compound eigenfrequencies (of the USC unit here). Due to that, we assume their presence to barely alter the results in the QST protocol and thus we neglect them. Sharper predictions will require the thorough study of the different phenomena imposing cutoffs in the system. As previously advanced, we have considered  ohmic environments characterized by the spectral density
\begin{align}
\mathcal{J}(\Delta_{kj})=\frac{\zeta\Delta_{kj}}{\varpi},
\label{EqC15}
\end{align}
where $\zeta$ and $\varpi$ correspond to $\gamma_\ell$ (bare relaxation rate) and $\omega_{q,\ell}$ (frequency) for the $\ell$-th CPB, or to $\kappa$ and $\omega_{r}$ for the LC oscillator, respectively. For the sake of simplicity  we assume in the following that $\gamma_{\ell}=\gamma$. Finally, we obtain the ME for the relaxation of both qubits and oscillator
\begin{align}
&\frac{d\rho(t)}{dt}=-{i}[{H}_\textrm{USC},{\rho}_S(t) ]+\sum_{j,k>j}(\Gamma^{jk}_\kappa+\sum_{\ell=1}^2\Gamma^{jk}_{\gamma,\ell})\\
&\bigg[[\bar{n}(\Delta_{kj},{T})+1]\mathcal{D}[\sigma_{jk}]\rho(t)+\bar{n}(\Delta_{kj},{T})\mathcal{D}[\sigma_{kj}]\rho(t)\bigg],\nonumber
\label{EqC16}
\end{align}
where
\begin{align}
\Gamma^{jk}_\kappa&=\frac{\kappa\Delta_{kj}}{{\omega}_{r}}|(X)_{jk}|^2,~\,\text{and}~~\,
\Gamma^{jk}_{\gamma,\ell}=\frac{\gamma\Delta_{kj}}{\omega_{q,\ell}} |(\sigma^{x}_\ell)_{jk}|^2
\end{align}
correspond to the dressed photon leakage rate for the resonator, and to the dressed qubit relaxation rate of $\ell$th CPB respectively, with $(\sigma^{x}_\ell)_{jk}=\bra{\psi_j}\sigma^x_\ell\ket{\psi_k}$, and $(X)_{jk}=\bra{\psi_j}{a}+{a}^\dagger\ket{\psi_k}$.

\subsection{Depolarizing channels and energy relaxation due to $\sigma^{z}$}
In this subsection, we derive a master equation considering the same free Hamiltonian for the system and environment, but changing the interaction term used in previous subsection for 
\begin{align}
H_{SB}=\sum_l\sigma_\ell^z(h_lb_l+h_l^* b_l^\dagger).
\end{align}
where the operator $\sigma_\ell^z$ corresponds to the $z$-Pauli matrix of the $\ell$-th CPB. This will provide us with a phenomenological description of long-time charge noise for the CPBs~\cite{Cottet2002,Koch2007}. Following the previous procedure, we write $H_{SB}$ in the interaction picture
\begin{equation}
\tilde{H}_{SB}(t)=\sum_{jkl}e^{i\Delta_{jk}t}Z_{jk}\sigma_{jk}(h_lb_le^{-i\nu_lt}+h_l^*b_l^\dagger e^{i\nu_lt}),
\end{equation}
where $\sigma_\ell^z=\sum_{jk}Z_{jk}\sigma_{jk}$, with $Z_{jk}=Z^*_{jk}=\bra{\psi_j}\sigma^z\ket{\psi_k}$. As in the previous section, the system operators are renamed as follow
\begin{align}
\tilde{S}_1&=\sum_{j,k>j}S_1e^{-i\Delta_{kj}t}, \tilde{S}_2=\sum_{j,k>j}S_2e^{i\Delta_{kj}t},
\label{EqC20}
\end{align}
where $S_1=Z_{jk}\sigma_{jk}$, $S_2=Z_{jk}\sigma_{kj}$, and ${S}_3=\sum_{j}Z_{jj}\sigma_{jj}$. Similarly for the bath, where its operators $\tilde{B}_1$ and $\tilde{B}_2$ were defined in Eq.~(\ref{EqC5}). After performing the Born-Markov approximation, with the secular one, we obtain 
\begin{widetext}
\begin{align}\nonumber
\frac{\Delta\tilde{\rho}(t)}{\Delta t} = & \frac{1}{\Delta t}\int_0^{\Delta t}d\tau \int_{t+\tau}^{t+\Delta t}dt_1\sum_{\Delta_{kj},\Delta_{k^\prime j^\prime}} e^{i(\Delta^\prime_{kj}-\Delta_{kj})t_1} e^{i\Delta_{kj}\tau} \bar{G}_{11}(\tau) \Big(S_1\tilde{\rho}(t)S_1^{\prime^\dagger} - S_1^{\prime^\dagger}S_1 \tilde{\rho}(t)\Big)\\
&+e^{i(\Delta_{kj}-\Delta^\prime_{kj})t_1} e^{-i\Delta_{kj}\tau}\bar{G}_{22}(\tau)\Big(S_2\tilde{\rho}(t)S_2^{\prime^\dagger}-S_2^{\prime^\dagger} S_2 \tilde{\rho}(t)\Big)+\bar{G}_{33}(\tau)\Big(S_3\tilde{\rho}(t)S_3-S_3 S_3 \tilde{\rho}(t)\Big)
+\rm{{H.c.}},
\label{EqC21}
\end{align}
\end{widetext}
where $\bar{G}_{\alpha\alpha}(\tau)$ $\{\alpha=1,2\}$ is the correlation function defined in Eq.~(\ref{cor}), $\bar{G}_{33}(\tau)=\lim_{\nu_\ell\rightarrow0}\sum_{\ell}e^{i\nu_\ell t}|h_\ell|^2\bar{n}(\nu_\ell,T)+e^{-i\nu_l \tau} |h_l|^2\big[\bar{n}(\nu_l,T)+1\big]$, and $S^\prime_1=Z_{j^\prime k^\prime}\sigma_{j^\prime k^\prime}$, $S^\prime_2=Z_{j^\prime k^\prime }\sigma_{k^\prime j^\prime}$. In this case, we cannot neglect the terms proportional to $S_{3}$ for the depolarizing channel spectral density. Therefore, this interaction will be reflected in the total master equation with both a pure depolarizing and a relaxation channels~\cite{Beaudoin2011}. 

In the Schr\"odinger picture, and after applying the secular approximation, we obtain  
\begin{align}
\frac{d{\rho}(t)}{dt} = & -{i}[H_\textrm{USC},{\rho}(t) ]\\\nonumber
&+\sum_{\ell=1}^2\sum_{j,k>j}\Gamma_{\phi,\ell}^{jk}\bigg[[\bar{n}(\Delta_{kj},T)+1]\mathcal{D}[\sigma_{jk}]{\rho}(t)\\\nonumber
&+\bar{n}(\Delta_{kj},T)\mathcal{D}[\sigma_{kj}]{\rho}(t)\bigg]+\sum_{\ell=1}^2\sum_{j}\Gamma_{\phi,\ell}^{jj}\mathcal{D}[\sigma_{jj}]{\rho}(t),
\end{align}
where
\begin{align}
\Gamma_{\phi,\ell}^{jk}&=&\frac{\gamma^{\textrm{CPB}}_\varphi \Delta_{kj}}{\omega_{q,\ell}}|Z_{jk}|^2,\quad
\Gamma_{\phi,\ell}^{jj}=\frac{\gamma^{\textrm{CPB}}_\varphi}{2\omega^i_q}\omega_T|Z_{jj}|^2,
\end{align}
are the dressed off-diagonal depolarizing channel rate producing energy relaxation, and dressed pure depolarizing channel rate of the $\ell$-th CPB. Here also, $\gamma^{\textrm{CPB}}_{\varphi}$ is the bare depolarizing channel rate of both CPBs, and the thermal frequency $\omega_T=k_BT$. 
\subsection{Complete master equation}
Finally, we describe the dissipative dynamics for the external transmons  with standard weak-coupling limit master equation terms~\cite{Breuer2002,Kryszewski2008,Koch2007}, obtaining for the complete model
\begin{align}\nonumber
\frac{d{\rho}(t)}{dt} = & -{i}[H,{\rho}(t) ]+\sum^{2}_{m=1}\big[\gamma_m \mathcal{D}[\tau^x_m]\rho(t)+\gamma_{\phi_m} \mathcal{D}[\tau^z_m]\rho(t)\big]\\\nonumber
&+\sum_{j,k>j}\bigg[\Gamma^{jk}_{\uparrow{\rm{eff}}}\mathcal{D}[\sigma_{jk}]\rho(t)+\Gamma^{jk}_{\downarrow{\rm{eff}}}\mathcal{D}[\sigma_{kj}]\rho(t)\bigg]\\
&+\sum_{\ell=1}^2\sum_{j}\Gamma_{\phi,\ell}^{jj}\mathcal{D}[\sigma_{jj}]{\rho}(t),
\label{C27}
\end{align}
where the Hamiltonian $H$ is given by Eq. (\ref{eqH}), $\gamma_m$ and $\gamma_{\phi_m}$ are the relaxation and depolarizing channel rates of the $m$-th external transmon. $\Gamma^{jk}_{\downarrow{\rm{eff}}}$, and $\Gamma^{jk}_{\uparrow{\rm{eff}}}$ correspond to the effective decay rates of the system defined as
\begin{align}\nonumber
\Gamma^{jk}_{\downarrow{\rm{eff}}}&= \bigg[\Gamma^{jk}_\kappa+\sum_{\ell=1}^2\bigg(\Gamma^{jk}_{\gamma,\ell}+\Gamma_{\phi,\ell}^{jk}\bigg)\bigg]\bigg[\bar{n}(\Delta_{kj},T)+1\bigg],\\
\Gamma^{jk}_{\uparrow{\rm{eff}}}&= \bigg[\Gamma^{jk}_\kappa+\sum_{\ell=1}^2\bigg(\Gamma^{jk}_{\gamma,\ell}+\Gamma_{\phi,\ell}^{jk}\bigg)\bigg]\bar{n}(\Delta_{kj},T).
\end{align}

\end{document}